\pdfoutput=1
\documentclass[11pt,letterpaper]{article}

\usepackage{lmodern}
\usepackage[margin=1in]{geometry} 
\usepackage{microtype}           
\usepackage{titling}             
\usepackage{amsmath,amssymb,bm}
\usepackage{cite}
\usepackage{siunitx}        
\usepackage{graphicx}
\usepackage{booktabs}
\usepackage[colorlinks=true,allcolors=blue,linktocpage]{hyperref}
\usepackage[labelfont=bf]{caption}
\usepackage[version=4]{mhchem}  
\usepackage{xspace}
\usepackage{mathtools}
\usepackage{color} 

\begin{document}

\title{
	Probing Neutron Skins with KDAR Neutrinos:
	From Coherent to Diffractive Elastic Neutrino--Nucleus Scattering
}

\author{
\parbox{0.98\textwidth}{\centering
Kyoungsu Heo$^{1}$,
Heesung Kwon$^{1}$,
Jaewon Kim$^{1}$,
Jubin Park$^{1}$\thanks{Corresponding author. E-mail: \href{mailto:honolov@ssu.ac.kr}{honolov@ssu.ac.kr}},
Myung-Ki Cheoun$^{1}$\thanks{Corresponding author. E-mail: \href{mailto:cheoun@ssu.ac.kr}{cheoun@ssu.ac.kr}},
Eunja Ha$^{2}$\thanks{E-mail: \href{mailto:ejaha@hanyang.ac.kr}{ejaha@hanyang.ac.kr}},
Kyung Kwang Joo$^{3}$\thanks{E-mail: \href{mailto:kkjoo@chonnam.ac.kr}{kkjoo@chonnam.ac.kr}}
}
\\[0.6em]
\parbox{0.98\textwidth}{\centering\small $^{1}$Department of Physics and Origin of Matter and Evolution of Galaxy (OMEG) Institute, Soongsil University, Seoul 06978, Korea}
\\
\parbox{0.98\textwidth}{\centering\small $^{2}$Department of Physics and Research Institute for Natural Science, Hanyang University, Seoul 04763, Korea}
\\
\parbox{0.98\textwidth}{\centering\small $^{3}$Center for Precision Neutrino Research (CPNR), Department of Physics, Chonnam National University, Yongbong-ro 77, Puk-gu, Gwangju 61186, Korea}
}
\date{February 23, 2026}

\maketitle

\begin{abstract}
We investigate coherent elastic neutrino--nucleus scattering (CE$\nu$NS)
	induced by pion--decay--at--rest ($\pi$DAR) and kaon--decay--at--rest (KDAR)
	neutrinos, with emphasis on the transition from strict coherence to the
	diffractive regime.
	Organizing CE$\nu$NS observables in terms of the dimensionless variable $qR$,
	we show that $\pi$DAR measurements remain confined to the near--coherent
	region for all nuclei, whereas KDAR neutrinos ($E_\nu=236$~MeV) extend the
	kinematics into $qR\gtrsim1$, where recoil spectra develop genuine
	shape sensitivity to the nuclear weak form factor.
	Using representative light, medium--mass, and heavy nuclei
	($^{12}$C, $^{40}$Ca, $^{48}$Ca, and $^{208}$Pb), we examine relevant cross sections and quantify the statistical
	sensitivity to the neutron skin thickness achievable at a JSNS$^2$--like
	facility.
	For a total exposure of 10~ton$\cdot$year and realistic KDAR fluences,
	projected $1\sigma$ sensitivities reach
	$\Delta R_{np}^{\,(1 \sigma)}$ $\simeq0.09$--$0.02$~fm for $^{48}$Ca and
	$\simeq0.07$--$0.02$~fm for $^{208}$Pb as the fluence increases.
	These sensitivities are competitive with, and complementary to,
	parity--violating electron--scattering measurements such as CREX and PREX,
	while relying on an electroweakly clean neutral--current probe with distinct
	systematic uncertainties.	
	Our results establish KDAR--based CE$\nu$NS as a quantitatively robust and
	complementary avenue for probing neutron skins and nuclear weak densities
	beyond the coherent limit.
\end{abstract}


\section{Introduction}
Coherent elastic neutrino--nucleus scattering (CE$\nu$NS) is a neutral--current
process in which a neutrino couples coherently to the weak charge of an entire
nucleus.
At sufficiently low momentum transfer, scattering amplitudes from individual
nucleons add in phase, and the cross section scales approximately with the square
of the weak charge, $Q_W^2$.
Because the proton contribution is strongly suppressed by the factor
$1-4\sin^2\theta_W$\footnote{
	At the low momentum transfers relevant for CE$\nu$NS,
	we adopt a low--energy effective weak mixing angle appropriate for
	neutral--current processes,
	$\sin^2\theta_W(Q^2 \simeq 0)=0.2386$,
	which incorporates electroweak radiative corrections, including the running of
	$\sin^2\theta_W$, $\gamma$--$Z$ mixing, and box--diagram contributions.
	This choice is suitable for momentum transfers
	$Q\sim\mathcal{O}(10$--$100)\,\mathrm{MeV}$.
	The value quoted at the $Z$ pole,
	$\sin^2\theta_W^{\overline{\mathrm{MS}}}(m_Z)\simeq0.231$,
	corresponds to a much higher energy scale and is therefore not directly
	applicable to CE$\nu$NS processes.
},
the weak charge is dominated by neutrons, yielding the characteristic scaling
$Q_W^2\simeq N^2$.
CE$\nu$NS therefore provides an electroweakly clean and theoretically controlled
probe of neutron distributions and neutron skins in nuclei.

Early accelerator measurements of CE$\nu$NS employed neutrinos from pion decay at
rest ($\pi$DAR), which produce a prompt monoenergetic $\nu_\mu$ at $29.8$~MeV via
$\pi^+\!\to\!\mu^+\nu_\mu$, followed by delayed $\bar{\nu}_\mu$ and $\nu_e$ from
$\mu^+$ decay.
As a result, existing accelerator--based CE$\nu$NS measurements probe neutrino
energies around $E_\nu\simeq30$~MeV.
In contrast, kaon decay at rest (KDAR) provides a monoenergetic $\nu_\mu$ with a
substantially higher energy of $236$~MeV through $K^+\!\to\!\mu^+\nu_\mu$.
This increase in neutrino energy extends the accessible momentum--transfer and
recoil ranges, opening sensitivity to nuclear weak--form--factor effects beyond
the strictly coherent regime.

Since the first observation of CE$\nu$NS with a CsI[Na] detector at the
stopped--pion source of the Spallation Neutron Source (SNS) in
2017~\cite{COHERENT2017_Sci}, experimental studies have expanded rapidly across
targets and neutrino sources.
The COHERENT collaboration reported the first spectral measurement on argon in
2021~\cite{COHERENT2021_LAr} and subsequently presented evidence for CE$\nu$NS on
germanium~\cite{COHERENT2025_Ge_evidence}.
At reactor experiments, the CONUS+ collaboration reported a $3.7\sigma$
observation of CE$\nu$NS in the sub--keV to keV recoil regime in
2025~\cite{CONUSplus2025_Nature}.
In parallel, large liquid--xenon detectors developed for dark--matter searches
have observed CE$\nu$NS induced by solar $^8$B neutrinos, with first signals
reported by XENONnT and PandaX--4T in
2024~\cite{XENONnT_2024_solarCEvNS,PANDAX4T_2024_solarCEvNS}.

Across these diverse experimental settings, CE$\nu$NS measurements enable
precision tests of the Standard Model at low momentum transfer, including
determinations of the weak mixing angle that complement atomic parity violation
and parity--violating electron scattering.
They also provide direct sensitivity to neutron distributions through the weak
form factor $F_W(q)$, allowing constraints on the neutron rms radius $R_n \equiv \sqrt{\langle r^2\rangle_n}$ and the neutron skin thickness
$\Delta R_{np} \equiv R_n - R_p
= \sqrt{\langle r^2\rangle_n}-\sqrt{\langle r^2\rangle_p}$,
where $R_p \equiv \sqrt{\langle r^2\rangle_p}$ is the proton rms radius, once sufficient statistics are
accumulated~\cite{NuclearStruct_CEvNS_Review,Cadeddu2019_PRC}.
Beyond nuclear structure, CE$\nu$NS offers sensitivity to nonstandard neutrino
interactions and other physics beyond the Standard Model through combined
rate--and--shape analyses~\cite{BSM_CEvsNS_Review}.

Despite this progress, the neutron skin thickness of medium and heavy nuclei
remains a major open question.
Recent parity--violating electron--scattering measurements report
$\Delta R_{np}(^{48}\mathrm{Ca}) = 0.121 \pm 0.026~\mathrm{fm}$ (CREX) and
$\Delta R_{np}(^{208}\mathrm{Pb}) = 0.283 \pm 0.071~\mathrm{fm}$ (PREX--II)
\cite{PREXII,CREX}, while hadronic probes yield partially overlapping but not
fully consistent results~\cite{RCNP1,RCNP2}.
The present experimental uncertainties, at the level of
$\sim0.03$--$0.07$~fm, map directly onto uncertainties in the density dependence
of the nuclear symmetry energy and the pressure of neutron--rich matter in
neutron stars~\cite{Miyatshu2025}.
An independent electroweak probe of the neutron density, subject to different
and largely orthogonal systematic uncertainties, is therefore of considerable
interest.

At low momentum transfer, the CE$\nu$NS differential cross section is given
by~\cite{Freedman:1974}
\begin{equation}
	\frac{d\sigma}{dT}
	=
	\frac{G_F^2 M}{4\pi}\,Q_W^2
	\left(1-\frac{MT}{2E_\nu^2}\right)
	F_W(q)^2,
	\qquad
	Q_W\simeq N-(1-4\sin^2\theta_W)Z ,
	\label{eq:dsdT}
\end{equation}
showing that all nuclear--structure dependence enters through the weak form
factor, the Fourier transform of the weak (primarily neutron) density.
Here $G_F$ is the Fermi constant, $E_\nu$ is the incident neutrino energy,
$T$ is the nuclear recoil kinetic energy, $M$ is the nuclear mass,
and $Z$ ($N$) is the proton (neutron) number.The momentum transfer $q$ is related to the recoil energy by
$q=\sqrt{2MT}/(\hbar c)$ (see Sec.~2).
Sensitivity to neutron radii and surface properties increases sharply as
measurements extend from the strictly coherent regime ($qR \lesssim 1$) into the
coherent--to--diffractive transition, where $F_W(q)$ departs from unity and
develops a weak oscillatory structure.

Most existing CE$\nu$NS measurements employing $\pi$DAR neutrinos remain confined
to the near--coherent regime for all experimentally relevant nuclei, rendering
recoil spectra only weakly sensitive to neutron--skin effects~\cite{NuclearStruct_CEvNS_Review,Cadeddu2019_PRC}.
In this work we show that KDAR neutrinos provide a qualitatively different
kinematic reach.
For medium and heavy nuclei, KDAR CE$\nu$NS accesses the transition region
$qR \gtrsim 1$ and, in the high--recoil tail, approaches the vicinity of the first
diffractive minimum.
Under JSNS$^2$ (J-PARC Sterile Neutrino Search at the J-PARC Spallation Neutron Source)--like benchmark conditions, this kinematic lever arm translates
into projected statistical sensitivities reaching $\Delta R_{np}^{\,(1\sigma)}\sim0.03$--$0.05$~fm for $^{48}$Ca and
$\sim0.02$--$0.03$~fm for $^{208}$Pb with a total exposure of
10~ton$\cdot$year, comparable to or improving upon current electroweak
constraints from CREX and PREX--II.
By combining analytic low--$q$ benchmarks with full weak--form--factor
calculations, we demonstrate how KDAR--induced recoil spectra can be exploited
to extract neutron--skin information in a controlled and quantitative manner.
These results establish KDAR CE$\nu$NS as a powerful and complementary
electroweak probe of neutron structure beyond the strictly coherent limit.

This paper is organized as follows.
Section~2 introduces the weak form factor formalism employed in this work and
summarizes the coherence condition and exact elastic kinematics used to map
recoil observables onto the momentum transfer.
Section~3 presents numerical results for representative targets, highlighting
how KDAR extends CE$\nu$NS from the near--coherent regime into the
coherent--to--diffractive transition and how this affects recoil spectra.
Section~4 develops a perturbative interpretation of neutron--skin sensitivity,
connects variations of the weak form factor to variations of $\Delta R_{np}$,
and presents both the kinematic constraints and the JSNS$^2$--based quantitative
sensitivity estimates (see Tables~\ref{tab:kdar_skin_summary} and \ref{tab:deltaR_sensitivity}).
Finally, section~5 summarizes our findings and discusses their implications for
future CE$\nu$NS cross section and neutron--skin measurements.

\section{Weak Form Factors, Coherence, and Kinematics in CE$\nu$NS}
The sensitivity of coherent elastic neutrino--nucleus scattering (CE$\nu$NS)
to nuclear structure is governed by the momentum transfer $\mathbf q$ relative
to the characteristic nuclear size $R$.
Physically, coherence reflects the inability of the exchanged neutral weak
boson to resolve spatial variations of the nuclear weak charge inside the
nucleus.
Throughout this work, we use the natural unit system, i.e., $\hbar = c = 1$.
Then the three--momentum transfer q can be expressed in terms of the reduced wavelength of the exchanged boson,
\begin{equation}
        \bar{\lambda}_{Z^0} \equiv \frac{\hbar c}{|\mathbf q|}
    = \frac{1}{q},
\end{equation}
such that coherent scattering occurs for $\bar{\lambda}_{Z^0} \gtrsim R$, equivalently $qR \lesssim 1$.
In this regime, scattering amplitudes from different regions of the nucleus
add coherently and the cross section exhibits the characteristic enhancement
proportional to $Q_W^2$.
As $qR$ increases beyond unity, spatial phases across the nuclear volume are
no longer aligned, the weak form factor departs from unity, and CE$\nu$NS
becomes sensitive to the internal structure of the nucleus.
This coherent--to--diffractive transition underlies the emergence of neutron
surface and neutron--skin sensitivity.

For $qR \gg 1$, the momentum transfer is sufficiently large enough to resolve
individual nucleons.
In this limit, scattering amplitudes add incoherently and elastic coherence is
lost, typically leading to inelastic processes.
The cross section then scales approximately with the number of scattering
centers rather than with the square of the weak charge.
Between these two extremes lies an intermediate elastic regime,
$qR \sim 1$, in which the nucleus recoils elastically but coherence is only
partial.
Here the event rate is suppressed by the nuclear weak form factor,
$|F_W(q)|<1$, and sensitivity to nuclear structure begins to develop.
This intermediate regime is of central importance for extracting neutron
radii and neutron skins from CE$\nu$NS data.

For spherically symmetric nuclei, the normalized nuclear weak form factor is
defined as
\begin{equation}
	F_W(q)
	=
	\frac{4\pi}{Q_W}
	\int_0^\infty r^2\, \rho_W(r)\, j_0(qr)\,dr,
	\qquad
	j_0(x)=\frac{\sin x}{x},
\end{equation}
where $\rho_W(r)$ denotes the weak--charge density and
\begin{equation}
	Q_W \equiv \int d^3r\,\rho_W(r)
	= N - (1-4\sin^2\theta_W)\,Z
\end{equation}
is the nuclear weak charge in the standard Freedman convention.
With this normalization, the weak form factor satisfies $F_W(0)=1$.
The weak--charge density can be decomposed into neutron and proton
contributions as
\begin{equation}
	\rho_W(r)
	=
	\rho_n(r) - (1-4\sin^2\theta_W)\,\rho_p(r),
\end{equation}
leading to
\begin{align}
	F_W(q)
	&=
	\frac{N\,F_n(q) - (1-4\sin^2\theta_W)\,Z\,F_p(q)}{Q_W},
	\notag\\
	F_{n,p}(q)
	&=
	\frac{1}{N\,(Z)}
	\int \rho_{n,p}(\mathbf r)\,
	\exp\!\left(i\frac{\mathbf q\cdot\mathbf r}{\hbar c}\right)\,d^3r .
\end{align}
Here $q\equiv|\mathbf q|$ is the corresponding wavenumber in fm$^{-1}$ used as the form-factor argument.
Because the proton contribution is strongly suppressed by the small factor
$(1-4\sin^2\theta_W)$, CE$\nu$NS observables are dominantly sensitive to the
neutron distribution.
As a result, the weak form factor closely tracks the neutron form factor,
$F_W(q)\simeq F_n(q)$, up to a small and well--controlled correction.

In this work we employ two standard parametrizations of the nuclear weak
density: the Helm model and the two--parameter Fermi (2pF) distribution.
These phenomenological descriptions capture the essential bulk and surface
features of neutron distributions relevant for CE$\nu$NS, while remaining
transparent and computationally efficient~\cite{Horo2012}.
More microscopic approaches, such as DRHBc or Skyrme--HFB--based calculations,
are beyond the scope of the present study.We left them as future work.

\subsection{Helm (top--hat $\otimes$ Gaussian) form factor}
The Helm form factor is given by
\begin{equation}
  F_{\rm Helm}(q)
  = \frac{3\,j_1(qR_0)}{qR_0}
  \exp\!\left[-\frac{(qs)^2}{2}\right],
  \label{eq:Helm}
\end{equation}
where $R_0$ is the diffraction radius that sets the location of the first
minimum of the form factor, and $s$ characterizes the surface thickness.
The corresponding weak--charge density reads~\cite{Horo2012}
\begin{equation}
	\rho_W(r)
	=
	\frac{3\,Q_W}{8\pi R_0^3}
	\left\{
	\operatorname{erf}\!\left(\frac{R_0 + r}{\sqrt{2}\,s}\right)
	-\operatorname{erf}\!\left(\frac{r - R_0}{\sqrt{2}\,s}\right)
	+ \sqrt{\frac{2}{\pi}}\,
	\frac{s}{r}
	\left[
	e^{-\frac{(r+R_0)^2}{2s^2}}
	- e^{-\frac{(r-R_0)^2}{2s^2}}
	\right]
	\right\}.
\end{equation}

The rms radius of the weak--charge distribution,
the weak radius $R_W$, is defined as
\begin{equation}
	R_W^2
	\equiv
	\frac{\int d^3r\, r^2\, \rho_W(r)}{Q_W},
\end{equation}
which evaluates to
\begin{equation}
	R_W^2
	=
	\frac{3}{5}\,\bigl(R_0^2 + 5s^2\bigr)
\end{equation}
for the Helm density.
Throughout this work, the characteristic radius appearing in the Helm
parametrization is taken as $R = r_0 A^{1/3}$.
This quantity should be regarded as an auxiliary nuclear--size scale rather
than the weak rms radius itself.
The diffraction radius $R_0$ is then fixed by $R_0^2 = R^2 - 5s^2$($R_{W}=\sqrt{3/5}R)$, ensuring a
consistent reproduction of the weak radius and the correct placement of
diffractive features in the weak form factor.

\subsection{Two--parameter Fermi (2pF / Woods--Saxon) form factor}
The two--parameter Fermi (2pF) distribution is defined as
\begin{equation}
	\rho(r)
	=
	\frac{\rho_0}{1+\exp\!\big[(r-c)/a\big]},
	\qquad
	F(q)
	=
	\frac{4\pi}{N}
	\int_0^\infty r^2 \rho(r)\, j_0(qr)\,dr .
	\label{eq:2pF}
\end{equation}
where $c$ and $a$ denote the half--density radius and surface diffuseness,
respectively.
In CE$\nu$NS applications, the weak form factor closely follows the neutron
form factor, with proton contributions suppressed by
$(1-4\sin^2\theta_W)$.
By construction, $\int \rho_n(\mathbf r)\,d^3r = N$, and the normalization
ensures $F_W(0)=1$.
Physically, $\rho_n(\mathbf r)$ represents the intrinsic spatial distribution
of neutrons, while $F_W(q)$ is the momentum--space quantity directly probed by
CE$\nu$NS data.

\subsection{Weak--charge densities from Helm and 2pF models}

\begin{figure}
	\centering
	\includegraphics[width=0.49\linewidth]{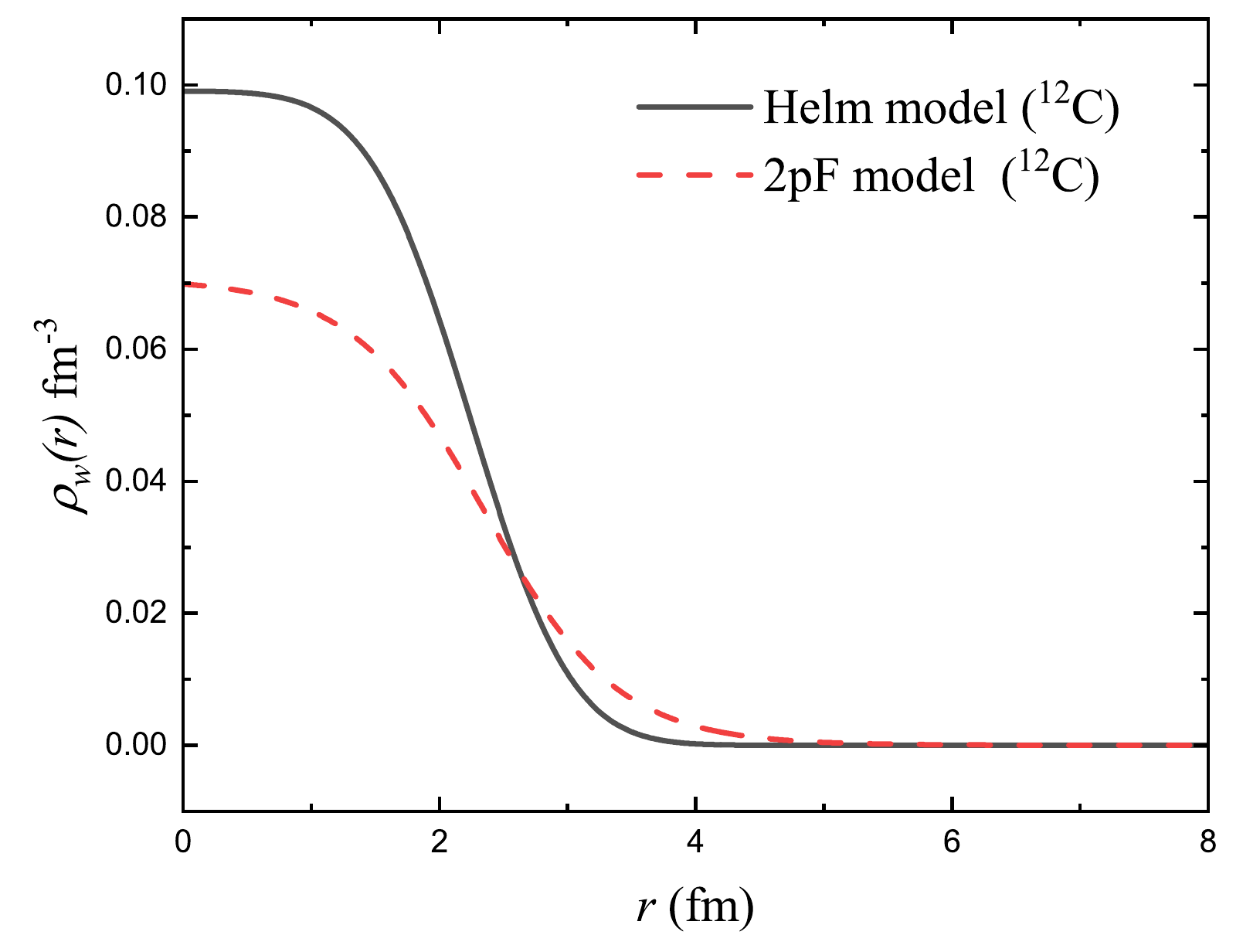}  
	\includegraphics[width=0.49\linewidth]{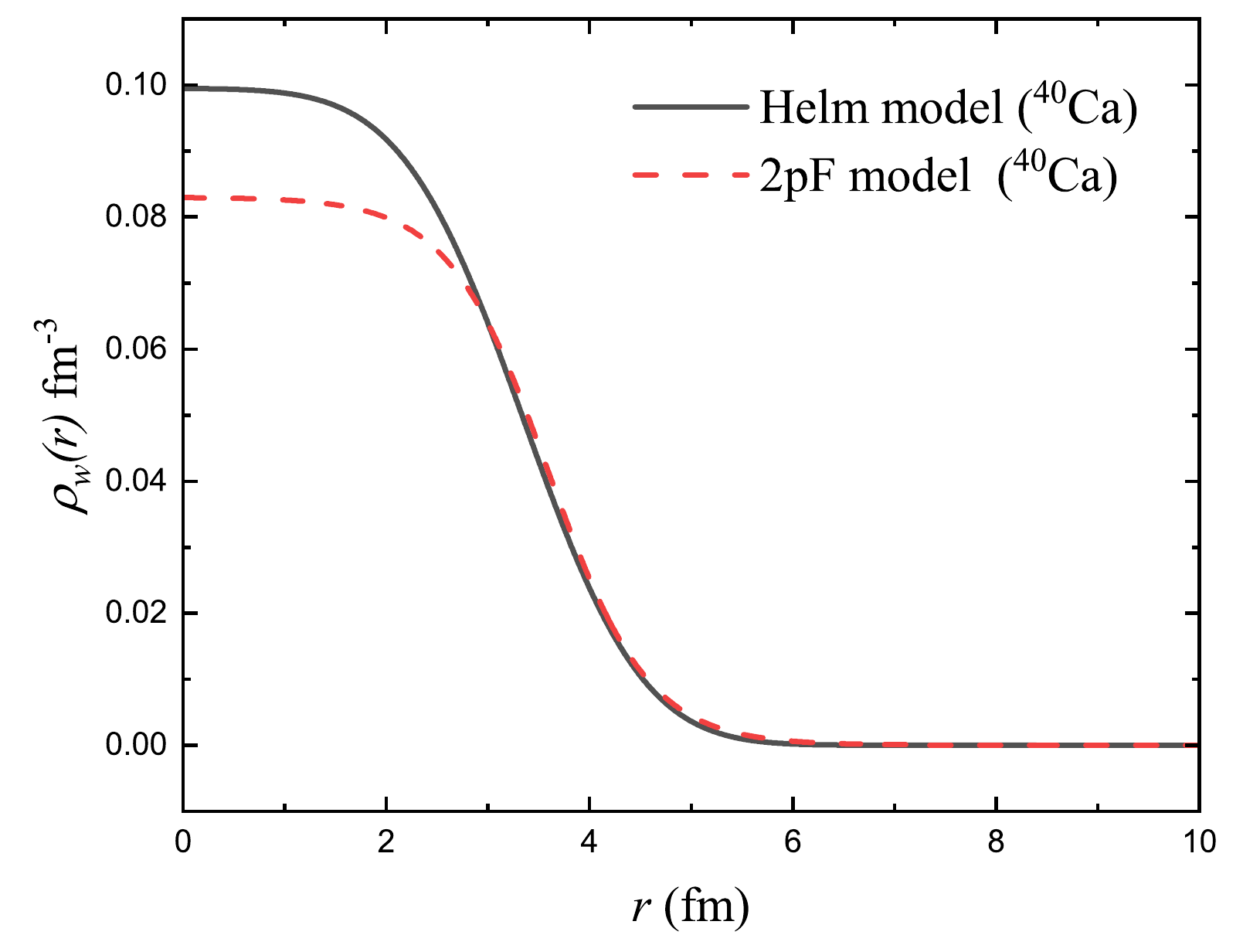}  
	\includegraphics[width=0.49\linewidth]{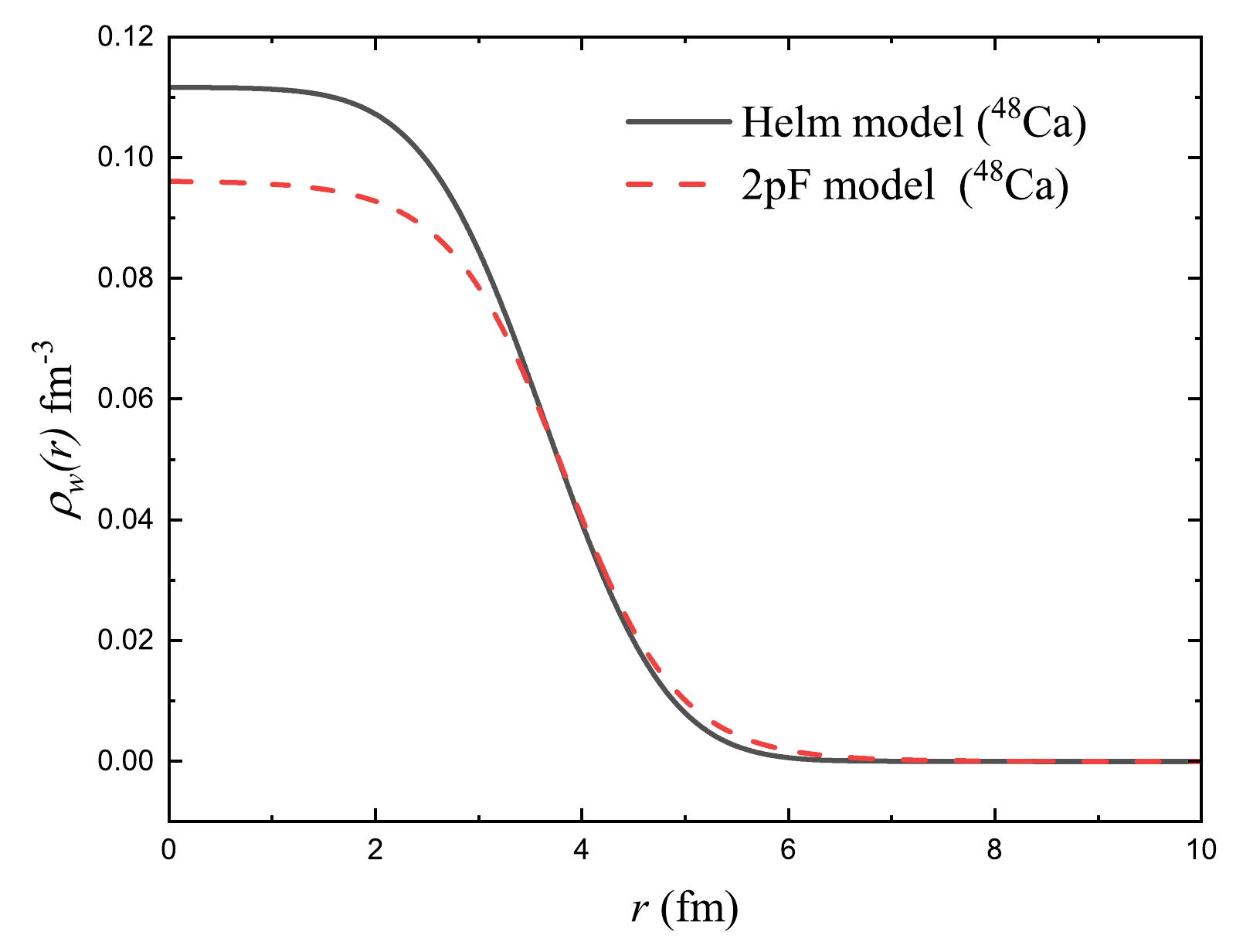}
	\includegraphics[width=0.49\linewidth]{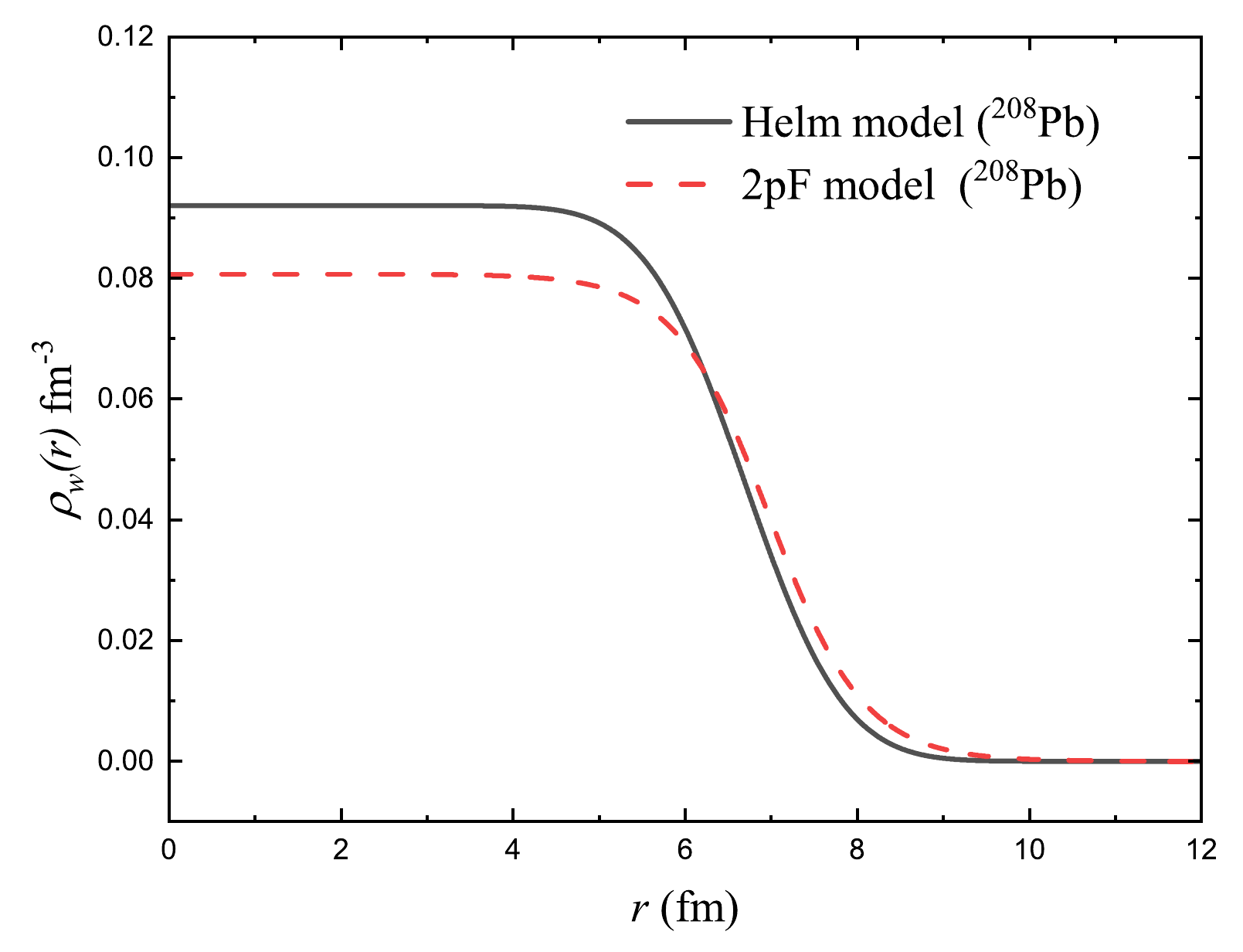}  
    \caption{
	Weak--charge densities $\rho_W(r)$ obtained with the Helm (solid)
	and two--parameter Fermi (2pF, dashed) parametrizations for
	$^{12}$C, $^{40}$Ca, $^{48}$Ca, and $^{208}$Pb.
	Each profile is normalized to the total weak charge $Q_W$
	(i.e., $\int d^3r\,\rho_W(r)=Q_W$).
	The parameter values used in each panel are summarized in Table~\ref{tab:presets}.}
	\label{fig:weak_densities}
\end{figure}

Figure~\ref{fig:weak_densities} compares the radial weak--charge densities
$\rho_W(r)$ obtained with the Helm and 2pF parametrizations for
$^{12}$C, $^{40}$Ca, $^{48}$Ca, and $^{208}$Pb.
All densities are normalized to the total weak charge $Q_W$ (i.e., $\int d^3r\,\rho_W(r)=Q_W$).
Representative parameter sets used for the Helm and 2pF densities are
summarized in Table~\ref{tab:presets}.

\begin{table}
	\centering
\begin{tabular}{l c c cc cc c}
	\toprule
	Isotope & $A$ & $r_0$ [fm]
	& \multicolumn{2}{c}{Helm}
	& \multicolumn{2}{c}{2pF}
	& Primary source \\
	\cmidrule(lr){4-5} \cmidrule(lr){6-7}
	& & 
	& $R_0$ [fm] & $s$ [fm]
	& $c$ [fm] & $a$ [fm]
	& \\
	\midrule
	C   & 12  & 1.20 & 2.40 & 0.60 & 2.355 & 0.52 & \cite{deVries1987,AngeliMarinova2013} \\
	Ca  & 40  & 1.20 & 3.58 & 0.90 & 3.593 & 0.49 & \cite{deVries1987,AngeliMarinova2013} \\
	Ca  & 48  & 1.20 & 3.87 & 0.90 & 3.670 & 0.55 & \cite{deVries1987,AngeliMarinova2013} \\
	Pb  & 208 & 1.20 & 6.82 & 0.90 & 6.624 & 0.55 & \cite{deVries1987,JonesBrown2014,AngeliMarinova2013} \\
	\bottomrule
\end{tabular}
\caption{
	Nuclear size parameters adopted to construct the weak--charge densities
	for the target nuclei considered in this work, using the Helm and
	two--parameter Fermi (2pF) form--factor parametrizations.
	In the Helm model, the effective diffraction radius $R_0$ and surface
	thickness $s$ are chosen to reproduce the corresponding weak rms radius,
	while the parameter $R_0$ is introduced solely as a conventional nuclear
	size scale.
	For the 2pF parametrization, $c$ and $a$ denote the half--density radius and
	surface diffuseness, respectively.}
    \label{tab:presets}
\end{table}
Across the four nuclei considered, the Helm and 2pF parametrizations reproduce
similar overall size scales and surface locations, but they distribute the weak charge
differently between the nuclear interior and the surface.
In particular, the Helm profile exhibits a flatter and systematically higher core density, whereas the 2pF form has a lower
central value and a more gradual decrease from the center.
The contrast is most pronounced for the light nucleus $^{12}$C and becomes less significant
for heavier systems; for $^{208}$Pb the two profiles are broadly similar over most of the
nuclear volume, with the remaining differences mainly visible in the surface tail at large $r$.
As long as the total weak charge and the low-order radial moments are similar,
differences in the core normalization are largely compensated, while surface/tail
differences drive the model dependence at intermediate $q$.

At larger momentum transfer, these surface differences translate into
corresponding variations of the weak form factor.
In the deeply coherent limit, $qR \ll 1$,
the Helm and 2pF models yield nearly identical predictions, while mild model dependence
sets in as $qR$ approaches $\mathcal{O}(1)$.
Near the first diffraction minimum at $qR \simeq 4.493$, they can
induce $\mathcal{O}(10\%)$ shape variations in the predicted recoil spectra for
medium--mass nuclei.
As the mass number increases, the relevant $qR$ values shift to higher
momenta, and the associated model dependence becomes less pronounced.
These trends provide a controlled estimate of the theoretical uncertainty
associated with weak--density modeling and motivate the use of both
parametrizations in the quantitative sensitivity analysis presented below.

\subsection{Exact elastic kinematics and experimental access to $qR$}
Experimentally, the momentum transfer is not measured directly.
Instead, CE$\nu$NS experiments observe the nuclear recoil energy $T$ and, in
some cases, the laboratory scattering angle.
For elastic neutrino--nucleus scattering, the recoil energy uniquely fixes
the momentum transfer and the scattering angle through
\begin{equation}
	q(T)=\sqrt{2MT},
	\qquad
	\theta(T)=\arccos\!\left(1-\frac{MT}{E_\nu(E_\nu-T)}\right),
\end{equation}
where $M$ is the nuclear mass and $E_\nu$ the incident neutrino energy.
This exact kinematical mapping provides a direct link between measured recoil
spectra and the momentum--space variables entering the weak form factor.

The maximum recoil energy and the accessible range in $qR$ increase with
neutrino energy.
Low--energy decay--at--rest sources predominantly probe the coherent regime,
with $q R \lesssim 1$ for light and medium--mass nuclei,
so that the weak form factor remains close to unity.
For heavy targets, $q R > 1$ is reached only near the kinematic endpoint.
Higher--energy neutrino sources extend the kinematic reach into the
coherent--to--diffractive transition and, for sufficiently heavy targets,
approach the vicinity of the first diffractive minimum at
$qR \simeq 4.493$.
These kinematical considerations clarify the complementary roles of different
neutrino sources and provide the foundation for the KDAR--based sensitivity
studies presented in the following sections.

\begin{table}[h]
	\centering
	\sisetup{round-mode=places,round-precision=0}
	\begin{tabular}{@{}lcccccc@{}}
		\toprule
		Target & $A$ & $R\,[\mathrm{fm}]$ & $E_\nu|_{qR=1}\,[\mathrm{MeV}]$ & $E_\nu|_{qR=4.493}\,[\mathrm{MeV}]$ & $T^{max}|_{qR=1}\,[\mathrm{keV}]$ & $T^{max}|_{qR=4.493}\,[\mathrm{MeV}]$ \\
		\midrule
		$^{12}\mathrm{C}$  & 12 & 2.747 & 36  & 161 &  $\sim$ 229 &  $\sim$ 4.529 \\
		$^{40}\mathrm{Ca}$ & 40 & 4.104 & 24  & 108 &  $\sim$ 23 &  $\sim$ 0.471 \\
		$^{48}\mathrm{Ca}$ & 48 & 4.361 & 23  & 102 &  $\sim$ 23 &  $\sim$ 0.460 \\
		$^{208}\mathrm{Pb}$ & 208 & 7.110 & 14  & 62 & $\sim$ 2 &  $\sim$ 0.040 \\

		\bottomrule
	\end{tabular}
\caption{
	Representative kinematic energy scales for the target nuclei considered in
	this work.
	Shown are the neutrino energies $E_\nu$ corresponding to the onset of coherence
	($q R \simeq 1$) and to the first diffraction minimum
	($q R \simeq 4.493$), together with the associated maximum nuclear recoil
	energies $T^{\max}$ at these kinematic points.
	The characteristic nuclear radius $R$ entering $q R$ is listed for each
	target.
}
	\label{tab:scales}
\end{table}


\subsection{PVES reference points (PREX-II and CREX) on the $qR$ axis}
\label{subsec:prex_crex_qR}

Parity--violating elastic electron scattering (PVES) experiments such as
PREX--II and CREX constrain the nuclear weak form factor at essentially a
\emph{single} acceptance--averaged momentum--transfer point
(i.e.\ one quoted reference value of $\langle Q^2\rangle$ or $q$)
\cite{PREXII,CREX}.
To compare those PVES constraints with the $qR$ systematics used throughout this
work, we translate the experimental reference point into the wavenumber
$q$ in fm$^{-1}$ and evaluate the corresponding dimensionless products $qR$
(and $qR_0$ for the Helm diffraction radius).

For a PVES reference point given in terms of the usual spacelike four--momentum
transfer $Q^2\equiv -q^\mu q_\mu$, we locate the point on our $qR$ axis by using
\begin{equation}
    q~[\mathrm{fm}^{-1}]
    \equiv
    \sqrt{\langle Q^2\rangle}.
    \label{eq:q_from_Q2_pves}
\end{equation}
Strictly speaking, PVES quotes the invariant $Q^2$ while in CE$\nu$NS kinematics we
use the three--momentum transfer $|\mathbf q|$; for elastic scattering the energy
transfer is small, so this identification is sufficient for the purpose of
placing PREX/CREX on the $qR$ axis.

For $qR$, we use the same nuclear--size convention as in Table~\ref{tab:scales},
$R=r_0A^{1/3}$ with $r_0=1.2~\mathrm{fm}$.
For the Helm model, we also quote $qR_0$ using
\begin{equation}
	R_0=\sqrt{R^2-5s^2},
	\label{eq:R0_from_R_s}
\end{equation}
with the same surface thickness $s$ values as in Table~\ref{tab:presets}.
The resulting PREX--II and CREX reference points lie in the intermediate window
$qR\simeq 3$--4, i.e.\ beyond the near--coherent region but still below the first
sharp--sphere zero at $qR\simeq 4.493$.

\begin{table}[t]
	\centering
	\caption{
		PVES momentum--transfer reference points for PREX--II and CREX and their
		location on the $qR$ axis.
		The dimensionless products $qR$ and $qR_0$ are evaluated with
		$R=r_0A^{1/3}$ ($r_0=1.2~\mathrm{fm}$) and the Helm diffraction radius
		$R_0=\sqrt{R^2-5s^2}$ using the same $s$ values as in Table~\ref{tab:presets}.
		Experimental reference points and $F_W(q)$ values are taken from
		Refs.~\cite{PREXII,CREX}.
	}
	\label{tab:prex_crex_qR}
	\begin{tabular}{@{}lcccccc@{}}
		\toprule
		Experiment & Target
		& Reference point
		& $q$ [fm$^{-1}$]
		& $qR$
		& $qR_0$
		& $F_W(q)$
		\\
		\midrule
		PREX--II & $^{208}\mathrm{Pb}$
		& $\langle Q^2\rangle = 0.00616~\mathrm{GeV}^2$
		& 0.398
		& 2.83
		& 2.71
		& $0.368\pm0.013$
		\\
		CREX & $^{48}\mathrm{Ca}$
		& $q = 0.8733~\mathrm{fm}^{-1}$
		& 0.8733
		& 3.81
		& 3.38
		& $0.1304\pm0.0052\pm0.0020$
		\\
		\bottomrule
	\end{tabular}
\end{table}
For visual reference, we overlay these PVES points on the corresponding weak form-factor curves in Fig.~\ref{fig:Form_factors}.
Because PVES constrains $F_W(q)$ at a single point in this intermediate-$qR$ window,
KDAR--driven CE$\nu$NS---which maps a continuous recoil range and can cover comparable
$qR$ values---provides a complementary electroweak probe of the same curvature region
of the weak form factor.


\section{From Coherence to Diffraction: Results for CE$\nu$NS Observables}
In this section we present numerical results for the nuclear weak form
factors and their impact on coherent elastic neutrino--nucleus scattering
(CE$\nu$NS) observables.
We first compare the Helm and two--parameter Fermi (2pF) parametrizations
at the level of the weak form factor itself, and then examine how these
features propagate into recoil spectra for KDAR
and pion--decay--at--rest ($\pi$DAR) neutrinos.
Throughout this section, the emphasis is placed on identifying the
kinematic regions in which CE$\nu$NS becomes sensitive to nuclear surface
structure beyond overall weak--charge normalization.

\subsection{Helm and 2pF weak form factors: from coherence to diffraction}
\begin{figure}
	\centering
	\includegraphics[width=0.42\linewidth]{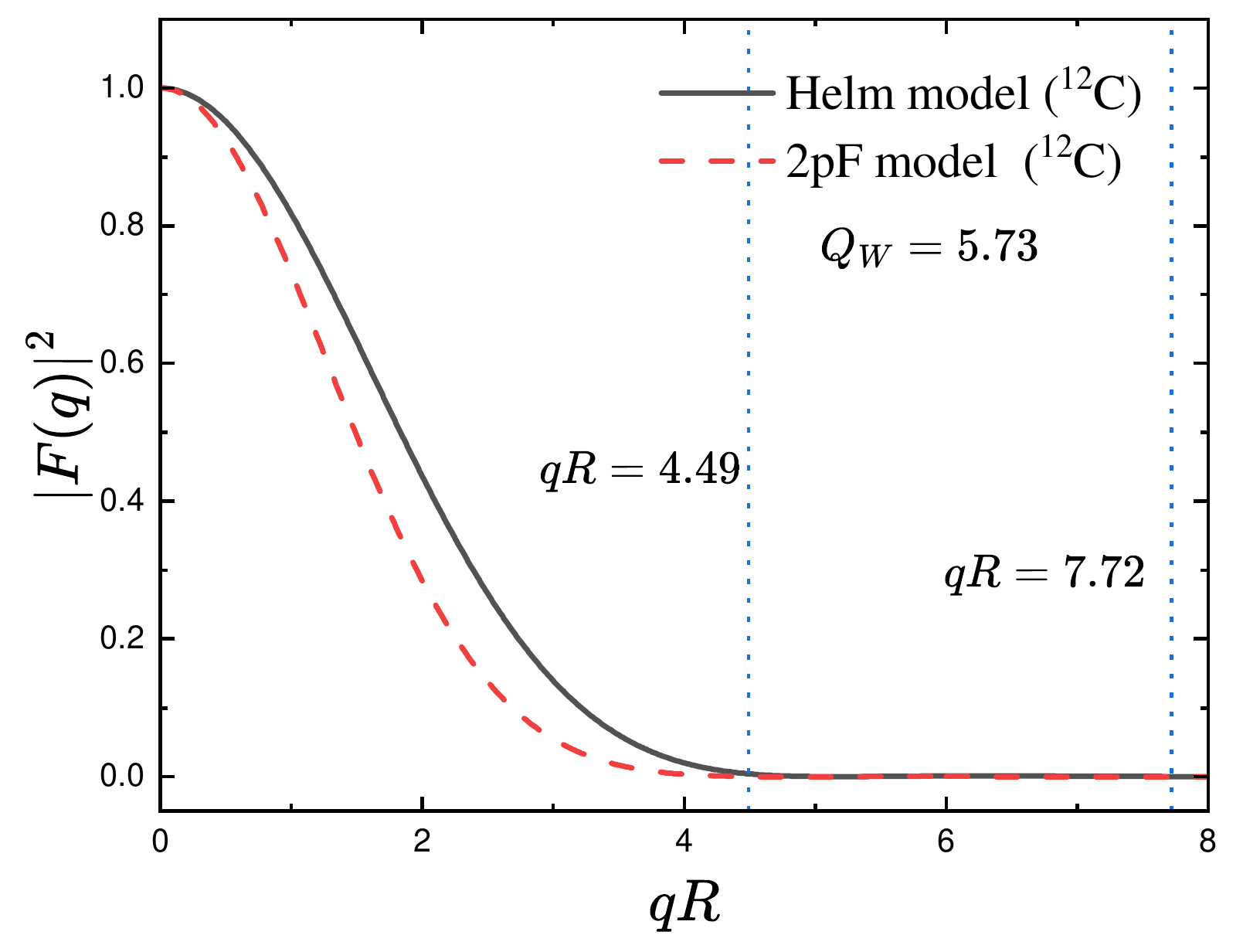}
    \includegraphics[width=0.42\linewidth]{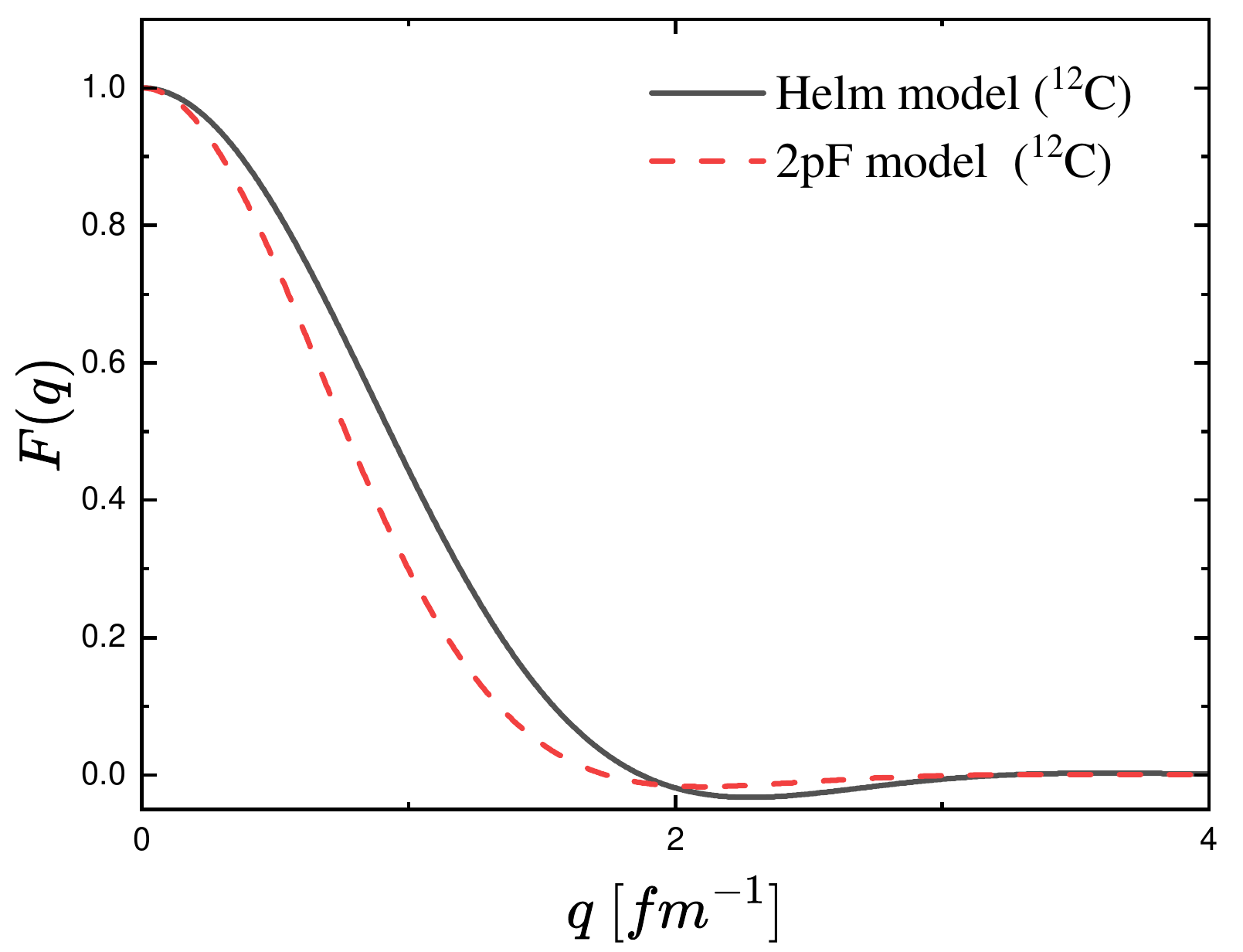}
	\includegraphics[width=0.42\linewidth]{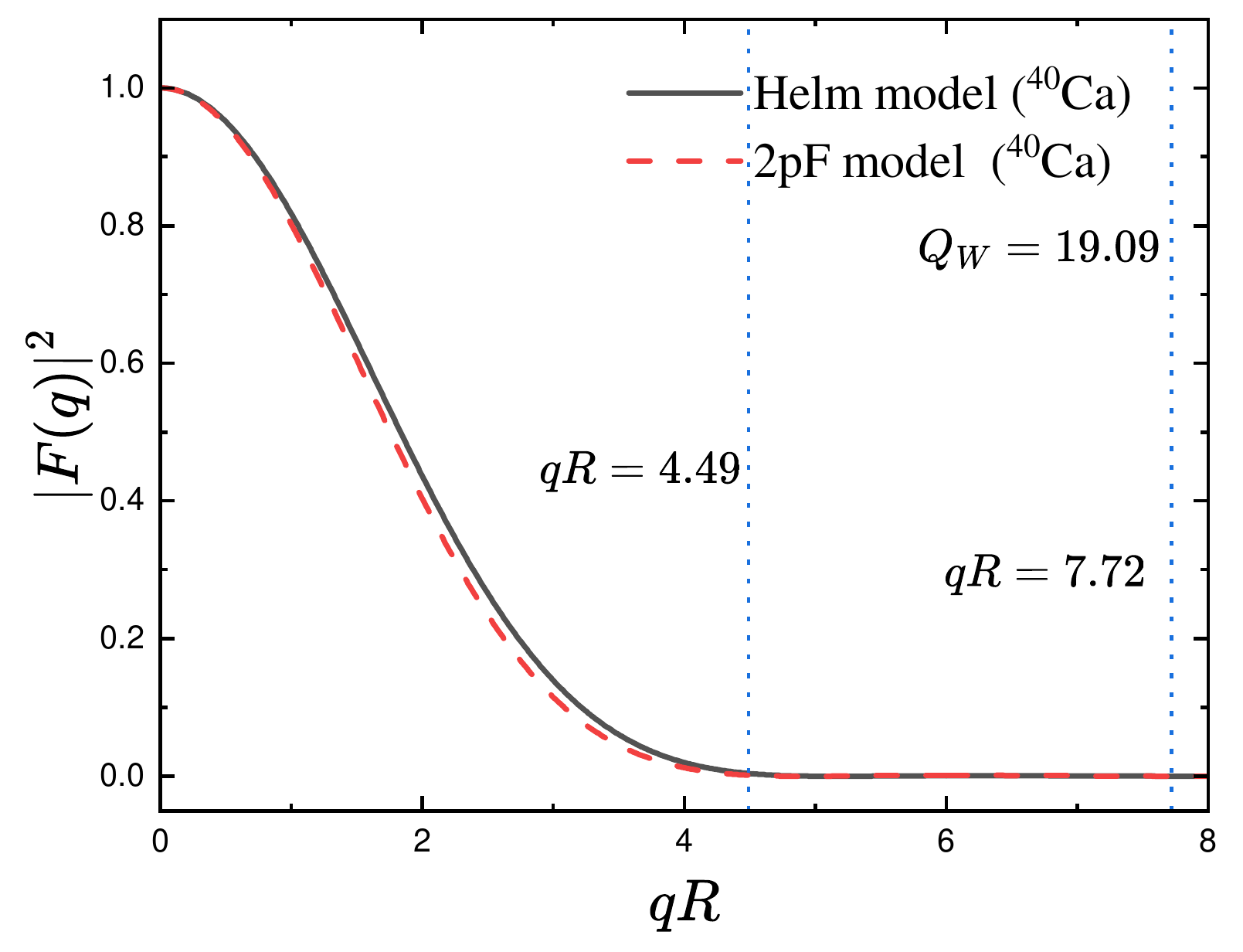}  
    \includegraphics[width=0.42\linewidth]{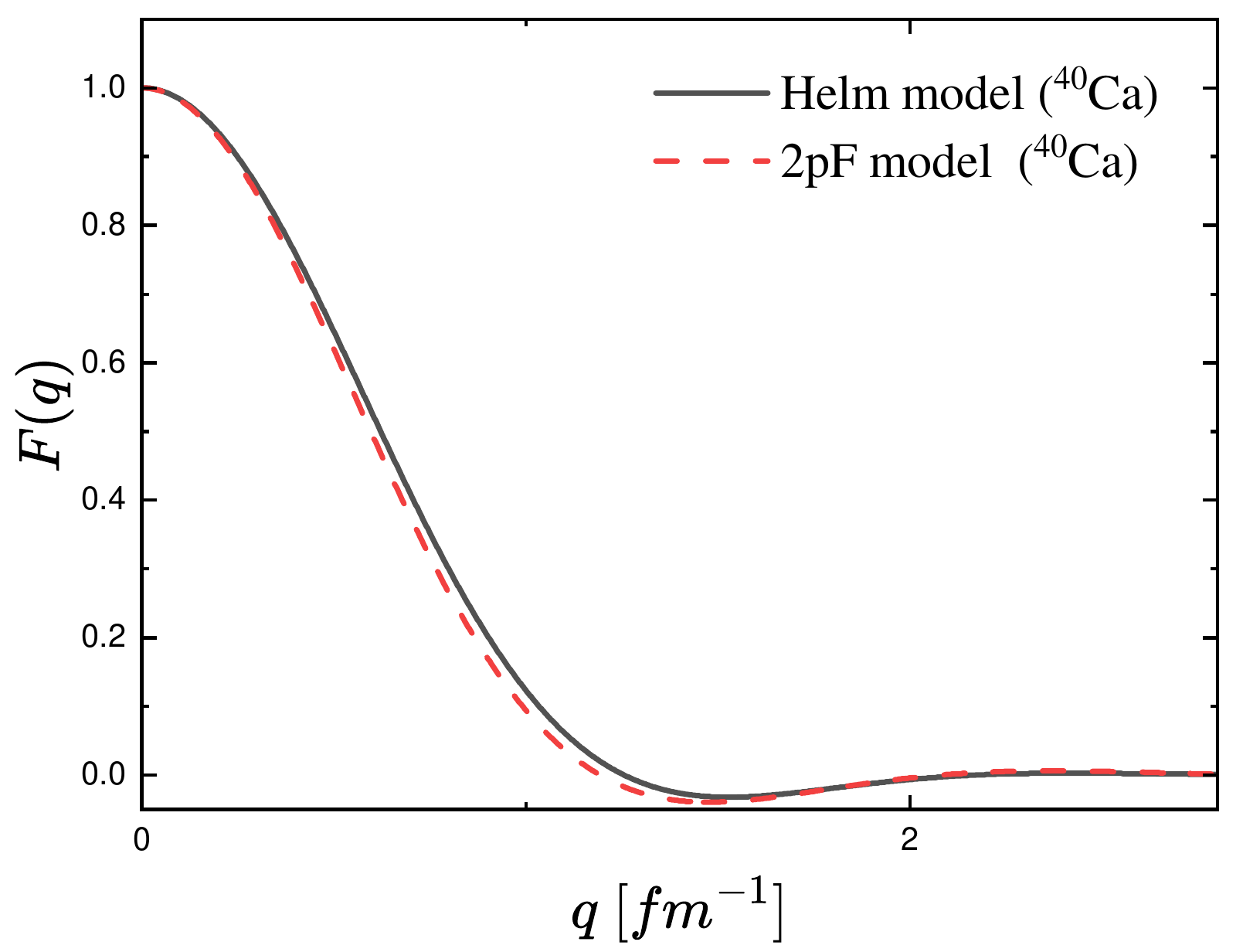}  
	\includegraphics[width=0.42\linewidth]{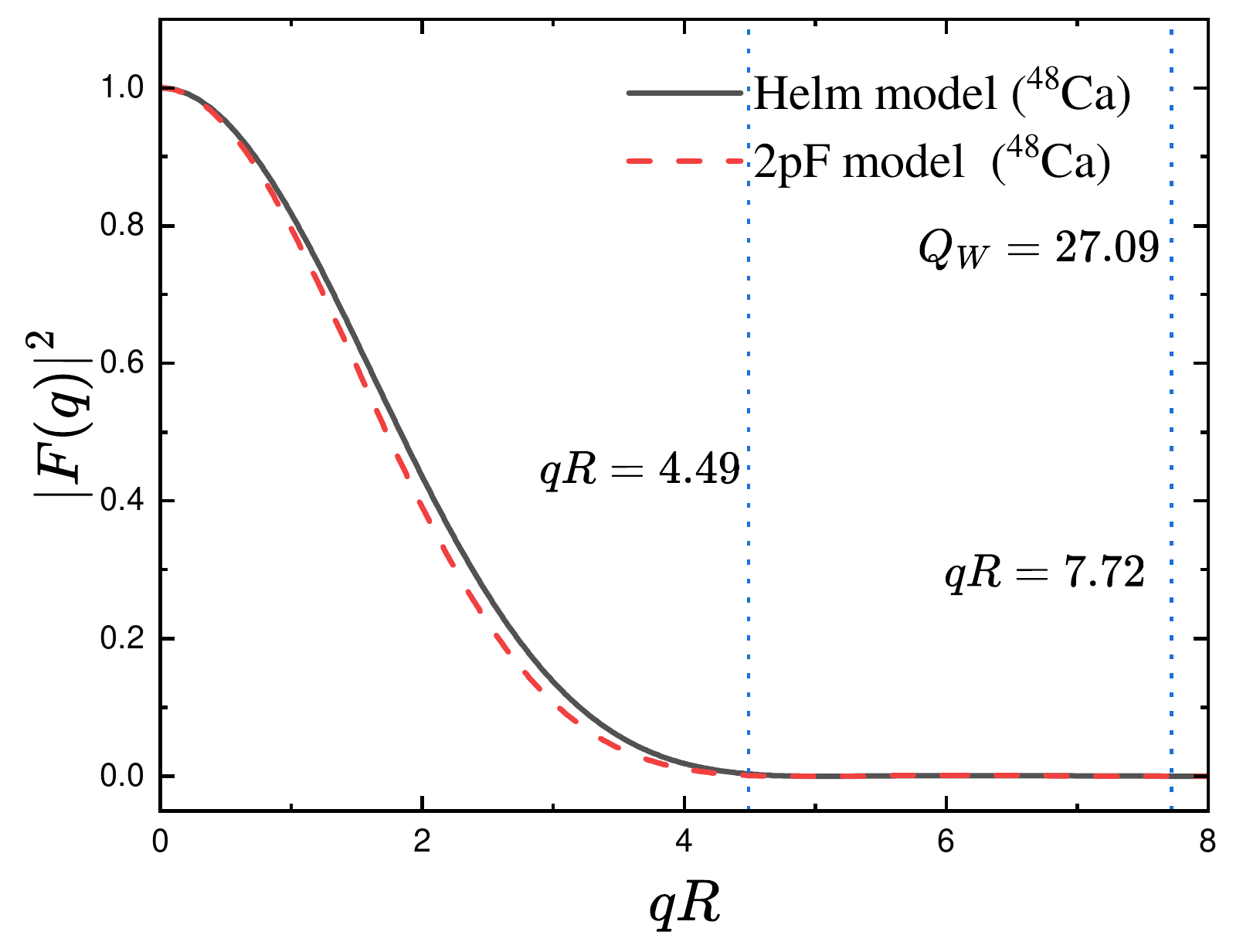}
    \includegraphics[width=0.42\linewidth]{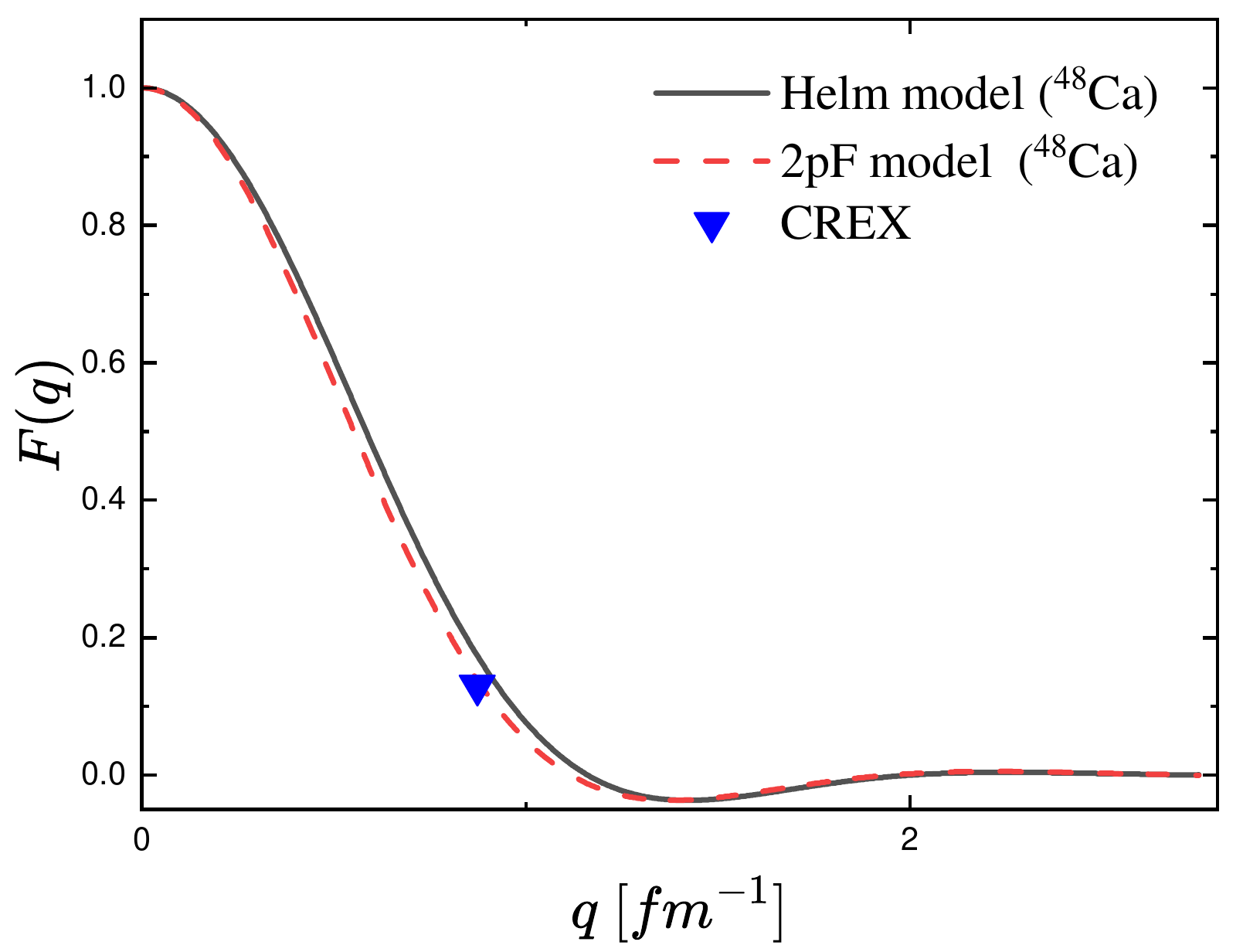}
	\includegraphics[width=0.42\linewidth]{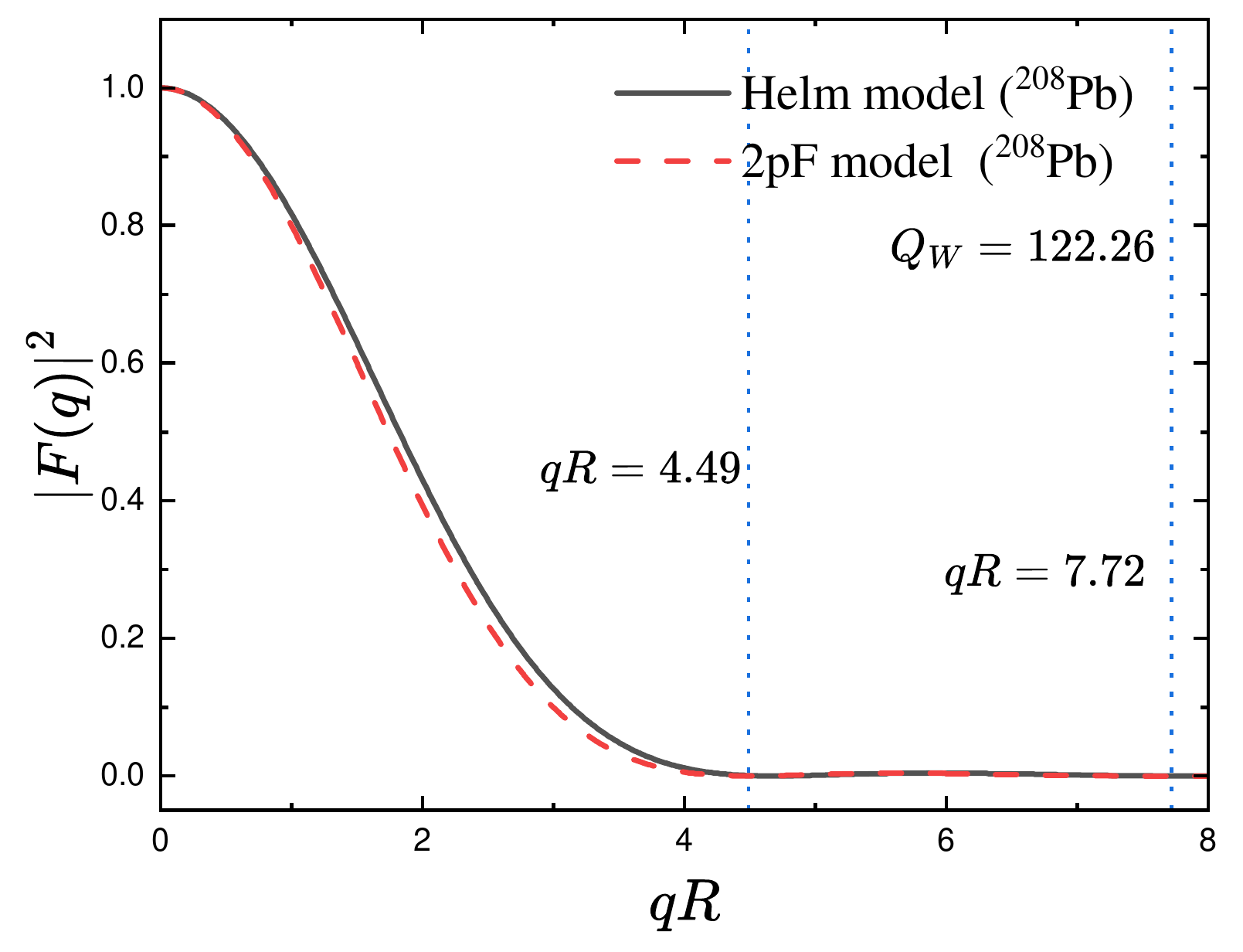}
    \includegraphics[width=0.42\linewidth]{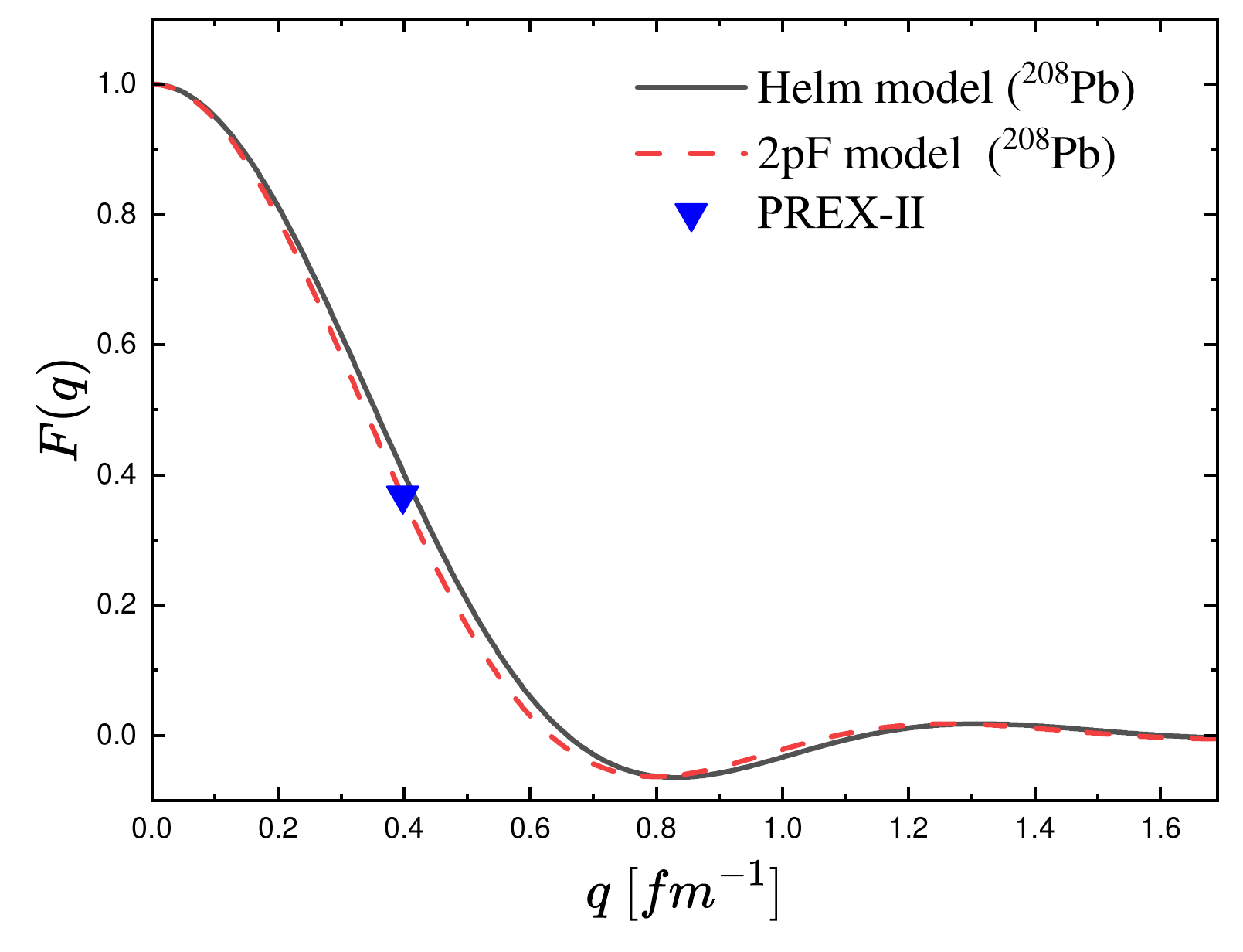}
	\caption{
		Weak form factors for $^{12}$C, $^{40}$Ca, $^{48}$Ca, and $^{208}$Pb
		calculated with the Helm (solid) and two--parameter Fermi (2pF, dashed)
		parametrizations.
		Left panels show $|F(q)|^2$ as a function of $qR$, and right panels show
		$F(q)$ as a function of $q$ (in fm$^{-1}$).
		Vertical dotted lines indicate the first and second diffraction minima
		at $qR\simeq4.493$ and $qR\simeq7.725$.
        Markers in the $^{48}$Ca and $^{208}$Pb right panels indicate the CREX and PREX-II measurements of $F_W(q)$ at their acceptance-averaged momentum transfers (Table~\ref{tab:prex_crex_qR}).}
	\label{fig:Form_factors}
\end{figure}

Figure~\ref{fig:Form_factors} compares the squared weak form factor
$F(q)^2$ as a function of the dimensionless variable $qR$ (left panels)
and the weak form factor $F(q)$ as a function of the physical momentum
transfer $q$ (right panels) for
$^{12}$C, $^{40}$Ca, $^{48}$Ca, and $^{208}$Pb.
Results obtained with the Helm parametrization are shown as solid curves,
while those based on the two--parameter Fermi (2pF) form are indicated by
dashed curves.
Vertical dotted lines mark the locations of the first and second
diffractive points of a sharp--sphere density at
$qR \simeq 4.493$ and $qR \simeq 7.725$.

\subsubsection{Behavior across kinematic regimes}

Several generic features are apparent.
By construction, both parametrizations satisfy $F(0)=1$, and they closely coincide in the
deeply coherent limit ($qR\ll 1$), where $|F(q)|^2$ remains near unity and model
dependence is minimal.
As $qR$ increases toward the edge of the coherent window, however, a non-negligible
suppression of $|F(q)|^2$ is already visible around $qR\sim\mathcal{O}(1)$, and the
onset of this suppression is mildly nucleus dependent (most noticeably for $^{12}$C).
Thus, $qR\sim 1$ should be regarded as a transition region where form--factor effects
begin to contribute, rather than a regime in which $F(q)$ can be treated as strictly unity.

As $qR$ increases further toward the first diffractive minimum (vertical dotted line),
both models exhibit the expected rapid damping and the emergence of oscillatory structure.
The Helm and 2pF curves remain broadly similar, but small differences develop in the
intermediate range and around the depth/location of the minima, reflecting the different
surface treatments encoded by $(R_0,s)$ versus $(c,a)$.
For light and medium--mass nuclei, the 2pF curve often lies slightly below the Helm curve
over part of the intermediate-$qR$ range, consistent with a smoother surface that
suppresses higher-momentum components; for $^{208}$Pb the two parametrizations are nearly
indistinguishable through the first oscillation, and any residual differences are pushed to
larger $qR$.

The mass dependence is further clarified in the right--hand panels,
which display $F(q)$ as a function of the wavenumber $q$ (in fm$^{-1}$).
The first diffractive minimum is determined by the condition
$q R \simeq 4.493$, implying
$q_{\text{node}} = 4.493/R$.
Thus $q_{\text{node}} \propto 1/R$, so that increasing $A$
(and therefore $R$) shifts the same diffractive feature
to smaller values of $q$ and compresses the coherent region
into a narrower interval in wavenumber.

\subsubsection{Implications for CE$\nu$NS observables}
The form--factor features shown in Fig.~\ref{fig:Form_factors} translate directly into
observable effects in CE$\nu$NS recoil spectra through
$d\sigma/dT \propto Q_W^2\,|F(q(T))|^2$.
In the deeply coherent region ($qR\ll 1$), both parametrizations yield nearly
identical predictions, so observables are governed primarily by the overall weak--charge
normalization.
As $qR$ approaches the edge of coherence ($qR\sim 1$) and extends into the
intermediate range ($qR\sim 2$--$5$), surface--model differences become visible and
can induce $\mathcal{O}(5$--$15)\%$ variations in the shoulder of the recoil spectrum for
light and medium--mass targets.
At still larger $qR$, the absolute rate is already strongly suppressed by
$|F(q)|^2$, so residual model differences have limited impact on the total event yield.

These trends indicate that light and medium--mass targets offer the
strongest lever arm for probing nuclear surface modeling and neutron
density systematics with CE$\nu$NS.
At the same time, combined analyses involving multiple nuclei can reduce
residual form--factor uncertainties in precision extractions, such as
constraints on neutron radii, neutron skins, or low--energy tests of the
weak mixing angle.

\subsection{Recoil kinematics for KDAR and $\pi$DAR neutrinos}

\begin{figure}[h] 
	\centering
	\includegraphics[width=0.49\linewidth]{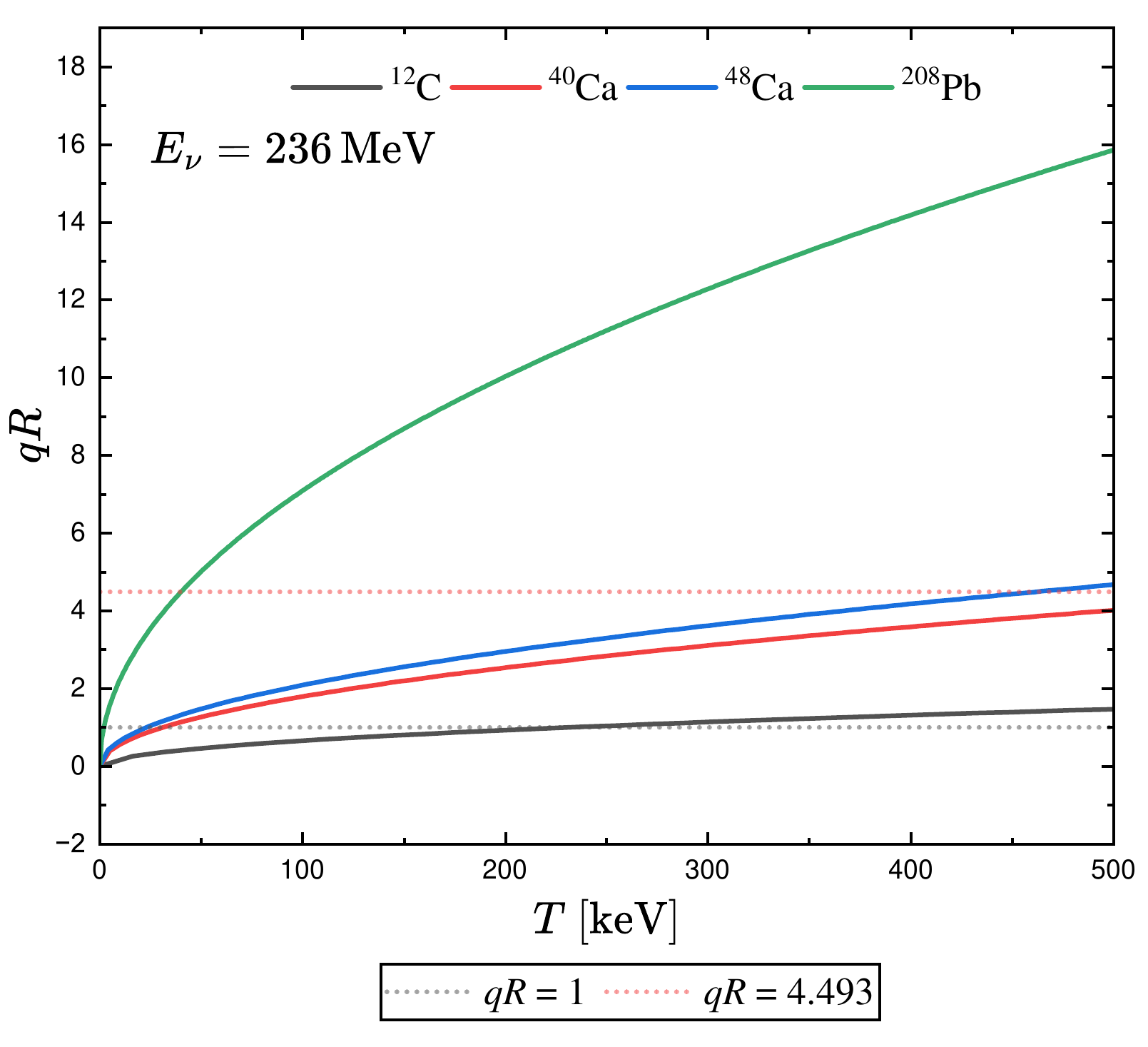}
    \includegraphics[width=0.49\linewidth]{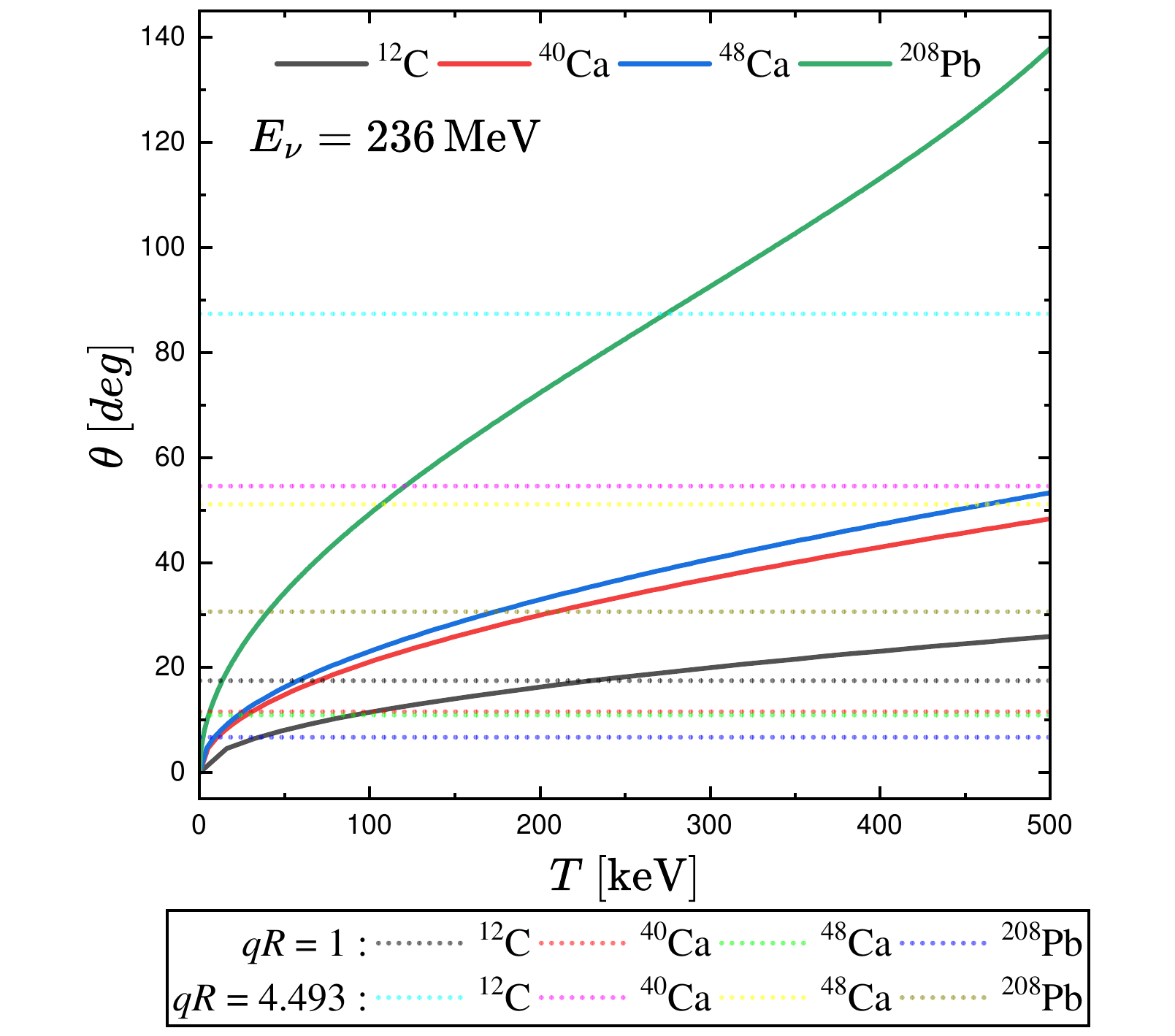}
	\includegraphics[width=0.49\linewidth]{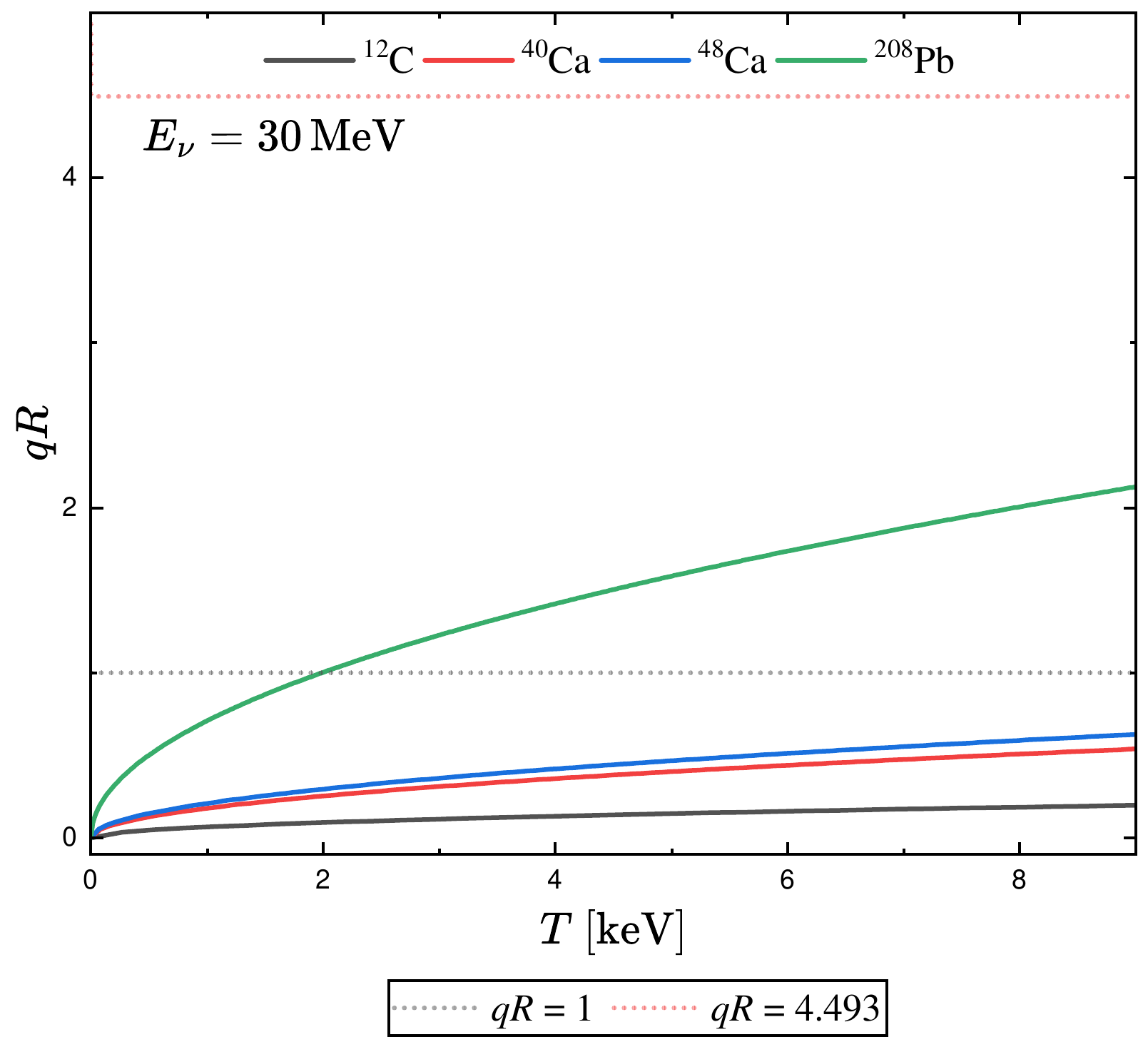} 
    \includegraphics[width=0.49\linewidth]{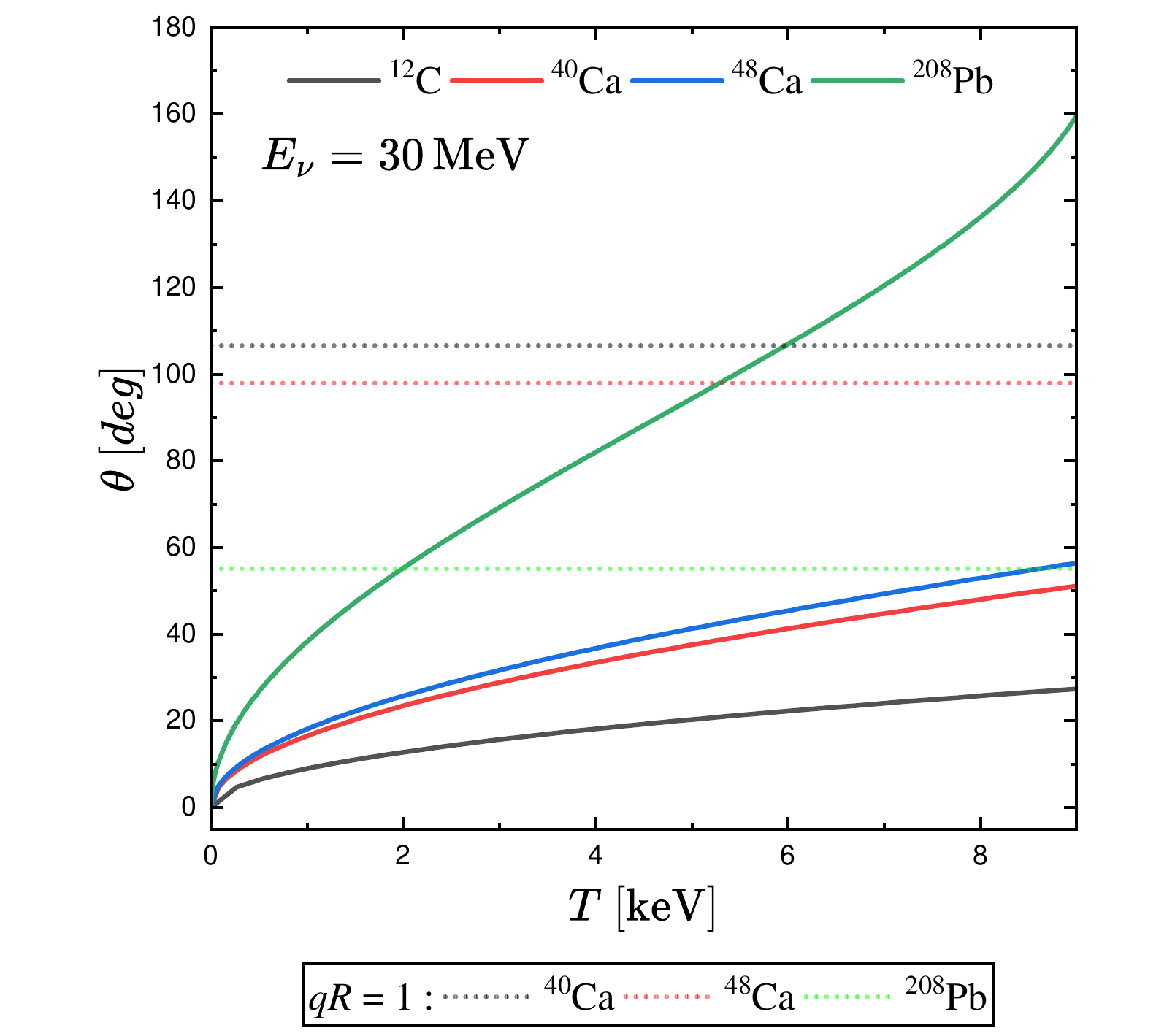} 
    \caption{
	Kinematical comparison of KDAR ($E_\nu=236$~MeV) and $\pi$DAR
	($E_\nu\simeq30$~MeV) neutrinos.
	Left panels show $qR$ as a function of the recoil energy $T$ for
	$^{12}$C, $^{40}$Ca, $^{48}$Ca, and $^{208}$Pb; horizontal dashed lines indicate
	$qR=1$ and $4.493$.
	Right panels show the corresponding laboratory scattering angle $\theta(T)$;
	for each target, dashed lines indicate the angles at which $qR=1$ and
	$4.493$ are reached.
	KDAR extends the kinematics to larger $T$ and $qR$, accessing the
	coherent--to--diffractive transition, while $\pi$DAR events are concentrated
	in the near--coherent region.}
	\label{fig:qR_kdar}
\end{figure}

Figure~\ref{fig:qR_kdar} illustrates how the substantially higher neutrino
energy of KDAR ($E_\nu=236$~MeV) modifies the recoil
kinematics relative to pion--decay--at--rest ($\pi$DAR,
$E_\nu\simeq30$~MeV) sources.
While elastic scattering kinematics uniquely relate recoil energy,
momentum transfer, and scattering angle, the higher KDAR energy shifts a
significant fraction of events into the region $qR\gtrsim1$ for light and
medium--mass nuclei.

In this kinematic domain, the recoil spectrum is no longer governed solely
by the weak--charge normalization but becomes directly sensitive to the
shape of the weak form factor and hence to the underlying neutron
distribution.
By contrast, $\pi$DAR kinematics are dominated by the near--coherent regime:
for light and medium--mass targets most events satisfy $qR\lesssim1$,
and even for heavy nuclei $qR>1$ is reached only near the kinematic endpoint.
Consequently, $F(q)$ stays close to unity over most of the $\pi$DAR recoil range,
and direct nuclear--structure sensitivity is intrinsically limited.
The higher recoil energies accessible with KDAR further improve
robustness against detector thresholds and place the characteristic
coherence and diffraction scales at experimentally accessible scattering
angles.

\subsection{CE$\nu$NS differential cross sections at KDAR and $\pi$DAR}

\begin{figure}
	\centering
	\includegraphics[width=0.49\linewidth]{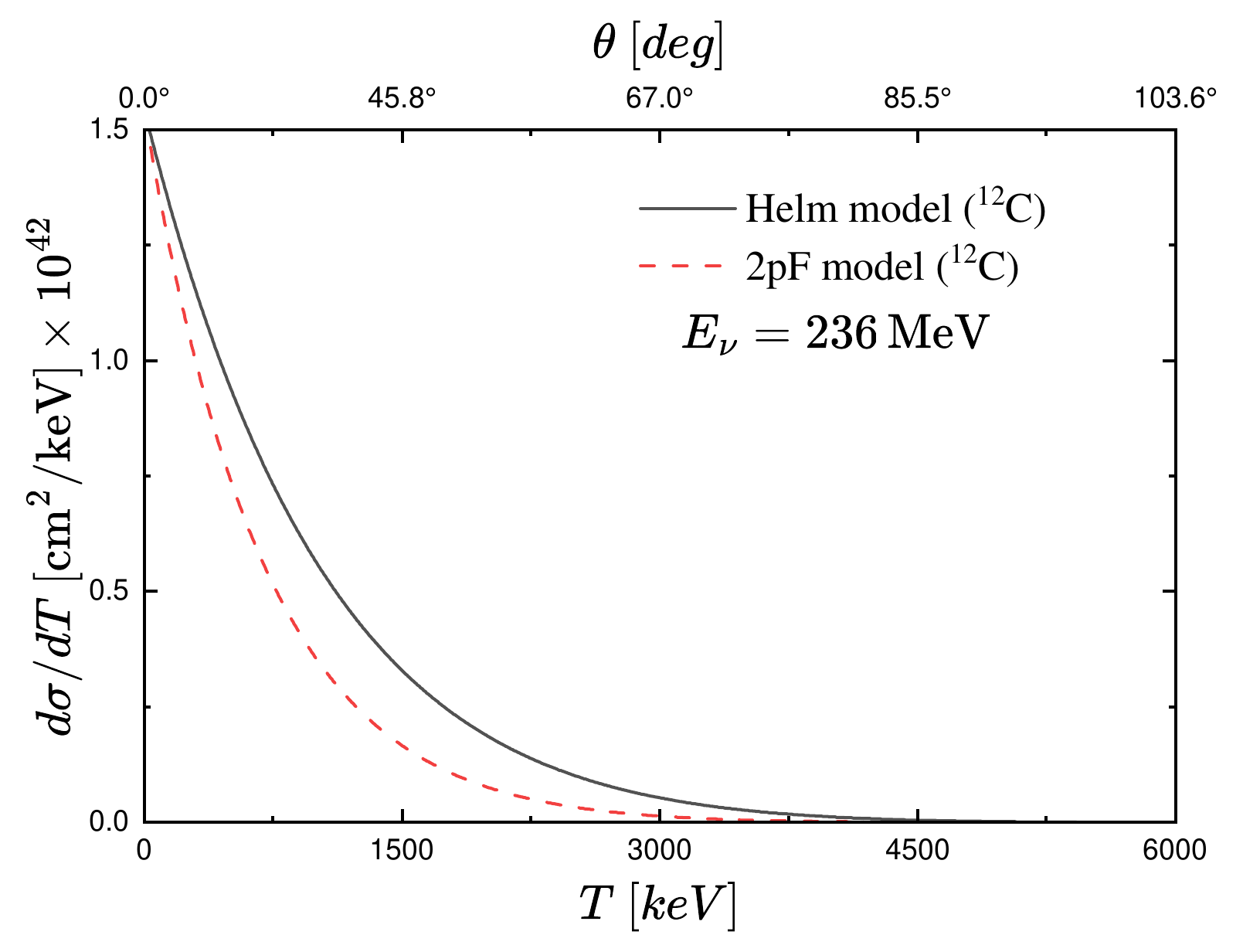}
	\includegraphics[width=0.49\linewidth]{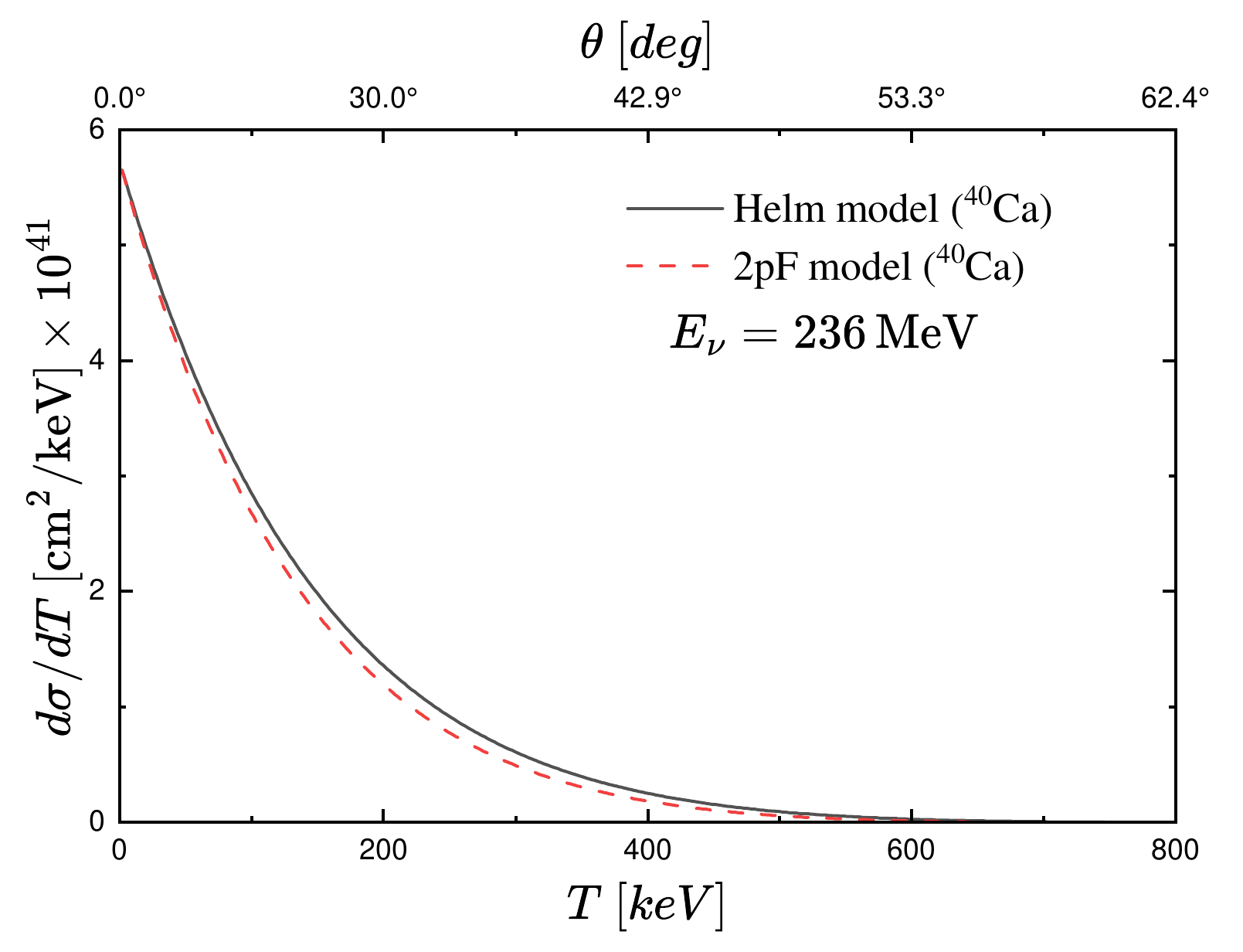}
	\includegraphics[width=0.49\linewidth]{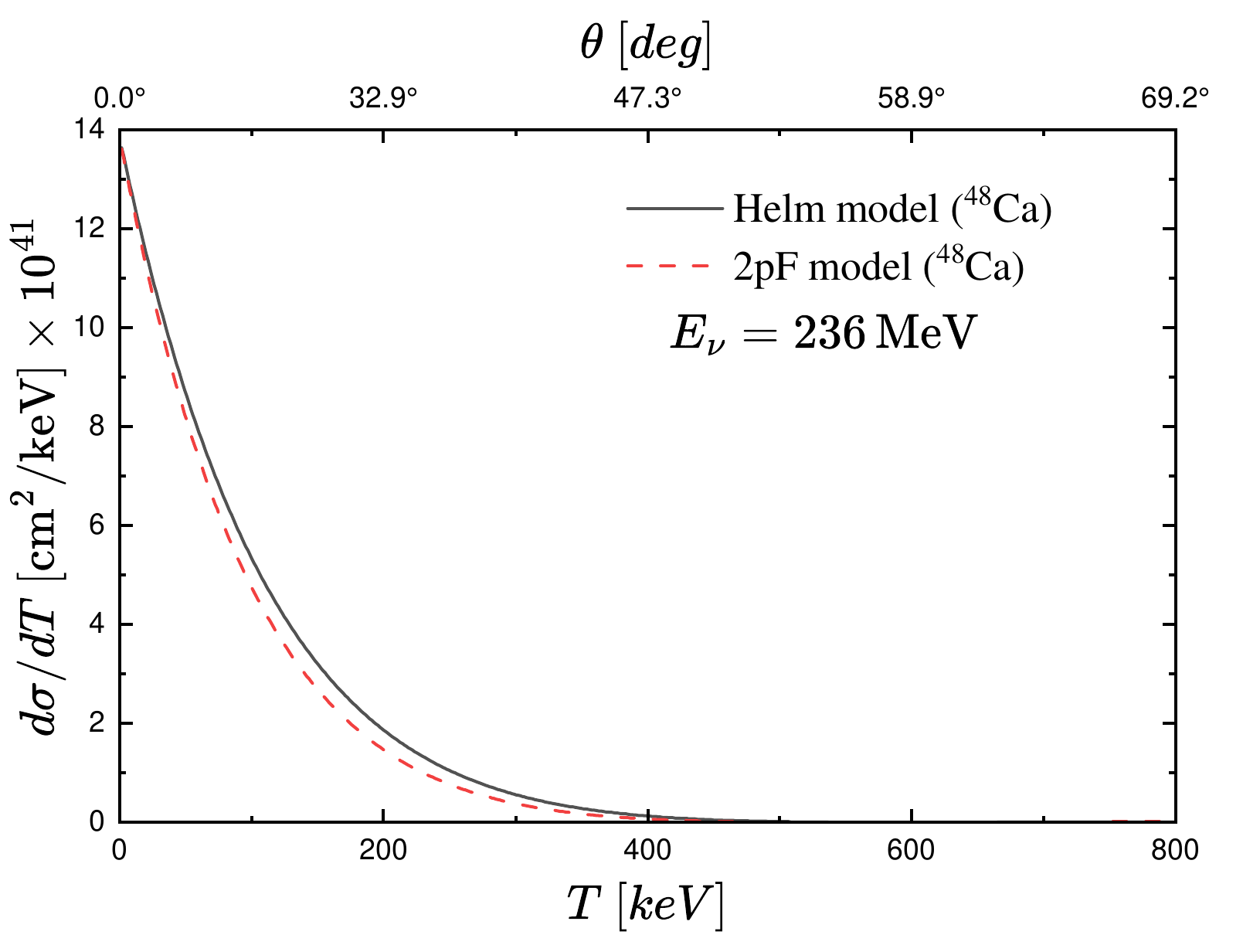}
	\includegraphics[width=0.49\linewidth]{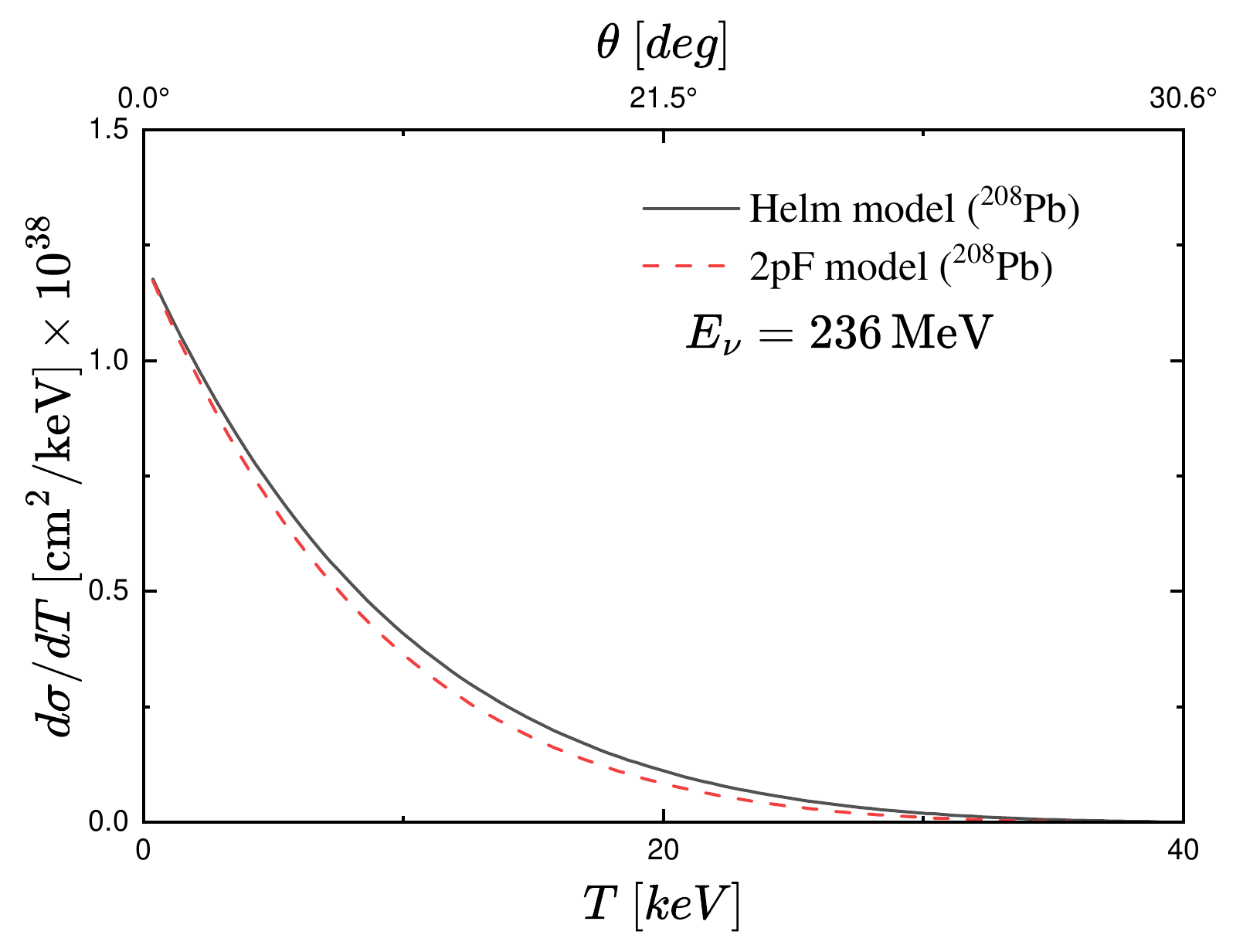}
\caption{
	CE$\nu$NS differential cross sections at KDAR ($E_\nu=236$~MeV) for
	$^{12}$C, $^{40}$Ca, $^{48}$Ca, and $^{208}$Pb, calculated with the
	Helm (solid) and two--parameter Fermi (2pF, dashed) weak form factors.
	Upper axes indicate the corresponding laboratory scattering angle.
	The spectra peak at low recoil energy and decrease rapidly with $T$;
	the overall normalization scales as $Q_W^2\propto N^2$, while
	model--dependent differences appear only at larger $T$, where
	form--factor suppression becomes significant.
}
\label{fig:dsdT KDAR}
\end{figure}

We now turn to the resulting CE$\nu$NS recoil spectra.
Figure~\ref{fig:dsdT KDAR} shows the predicted coherent elastic
neutrino--nucleus differential cross sections for
$^{12}$C, $^{40}$Ca, $^{48}$Ca, and $^{208}$Pb at
$E_\nu=236~\mathrm{MeV}$, computed using the Helm and two--parameter Fermi
(2pF) parametrizations of the weak form factor.
All spectra are evaluated using the leading--order CE$\nu$NS differential
cross section introduced in Eq.~(\ref{eq:dsdT}), with the momentum transfer
related to the recoil energy by $q=\sqrt{2MT}$ and $M\simeq A\,m_u$,
where $A$ is the mass number and $m_u$ is the atomic mass unit.
The upper axis in each panel maps the recoil energy $T$ to the corresponding
laboratory scattering angle, allowing spectral features to be interpreted
equivalently in energy or angle space.

\subsubsection{Global trends}
For all targets, the recoil spectrum is sharply peaked at low recoil
energies and decreases rapidly with increasing $T$.
The overall normalization follows the expected scaling
$Q_W^2\propto N^2$, explaining why $^{208}$Pb exhibits the largest
differential rate at a given recoil energy despite its smaller kinematic
endpoint.
Lighter nuclei display broader recoil distributions but with reduced
overall normalization.

\subsubsection{Form--factor dependence}
At very low recoil energies, where $qR \ll 1$, the Helm and 2pF parametrizations
yield indistinguishable spectra because $|F(q)|^2$ remains close to unity.
Differences emerge as $qR$ approaches and exceeds unity, where the two
models differ slightly in surface curvature and in the depth and location
of the first diffractive minimum.
At KDAR energies, these effects lead to modest but non--negligible
deviations in $d\sigma/dT$ for light and medium--mass nuclei at larger
recoil energies, while remaining small in the near--threshold region
that dominates the total event rate.

\subsubsection{Implications for KDAR versus $\pi$DAR}

\begin{figure}
	\centering
    \includegraphics[width=0.49\linewidth]{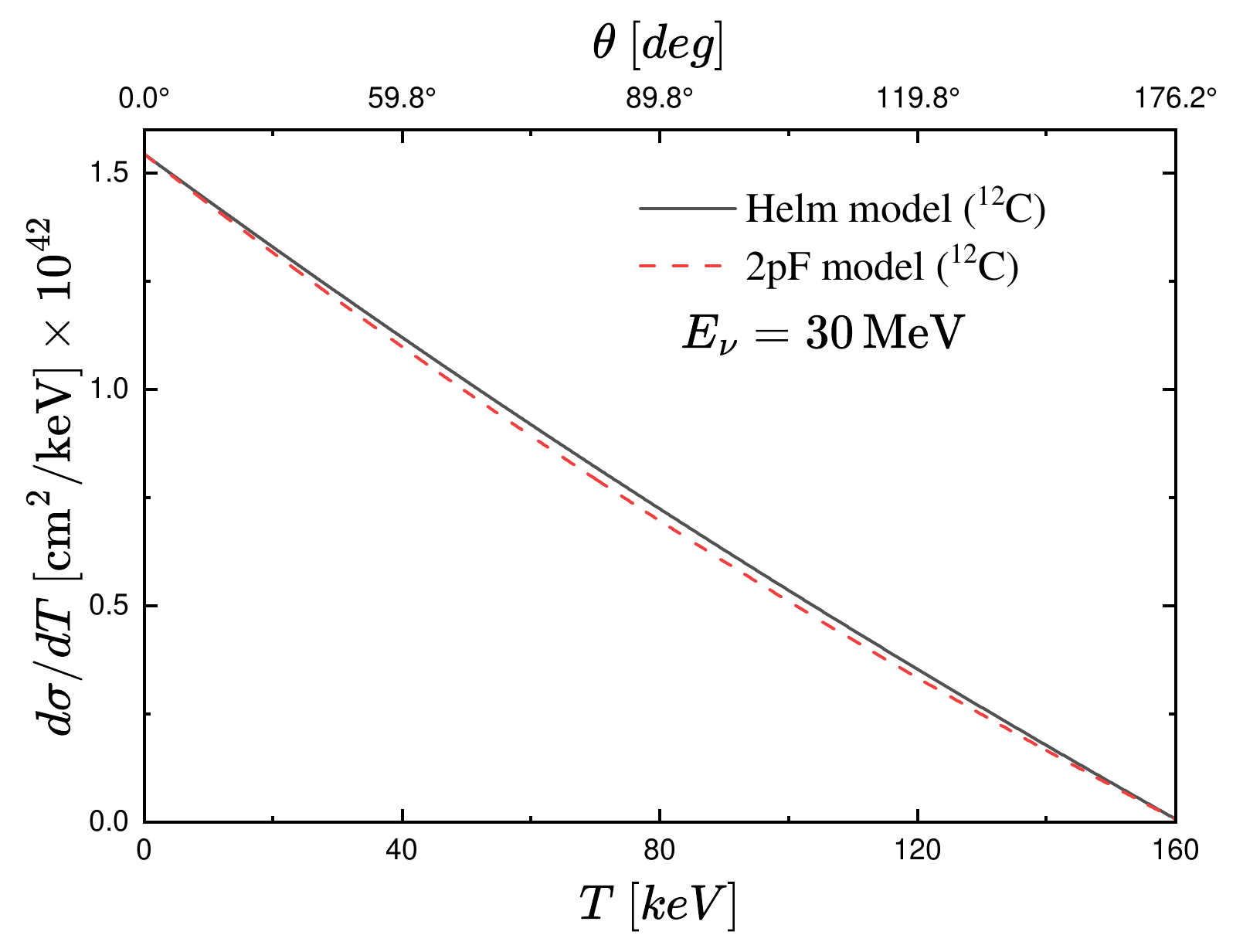}
	\includegraphics[width=0.49\linewidth]{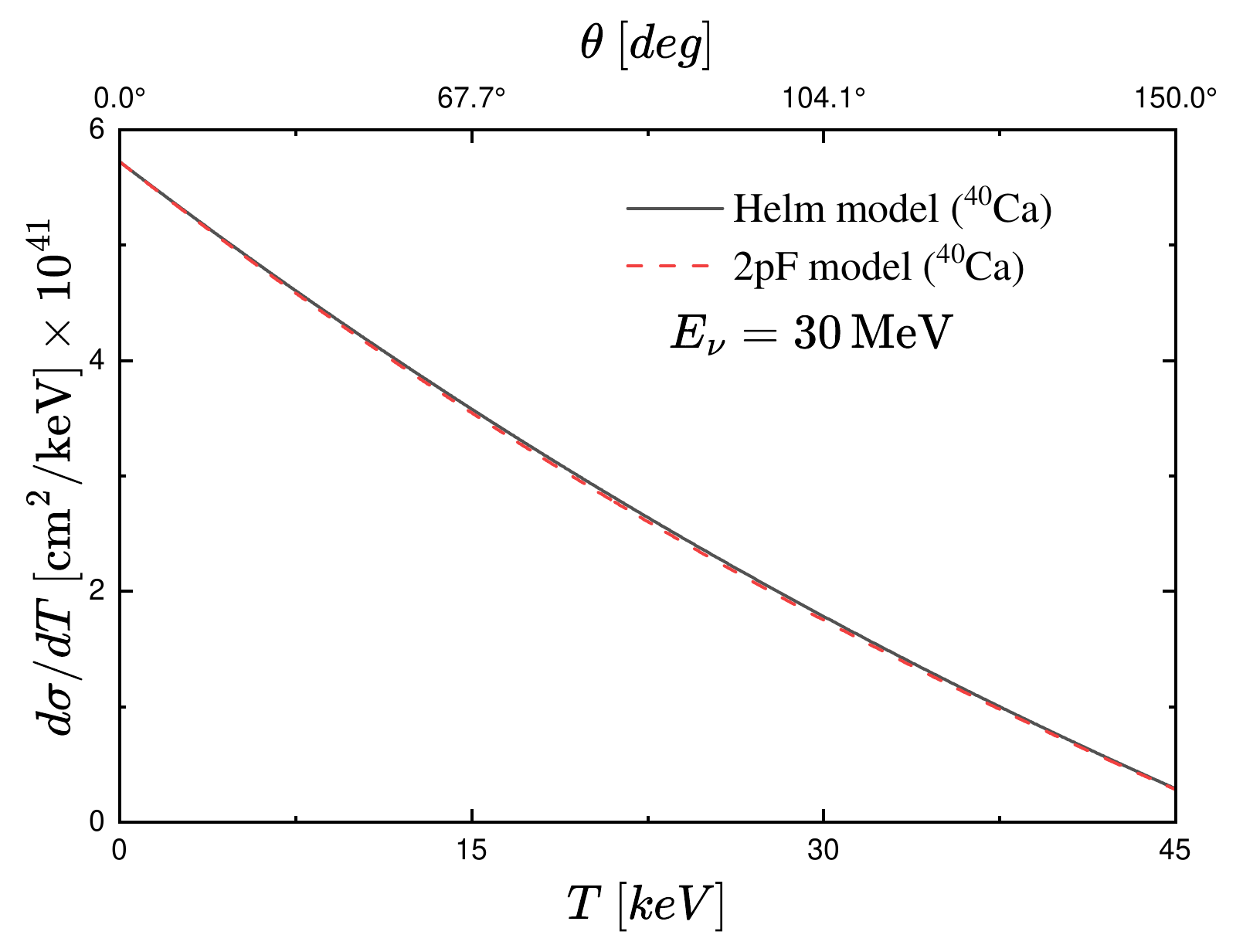}
	\includegraphics[width=0.49\linewidth]{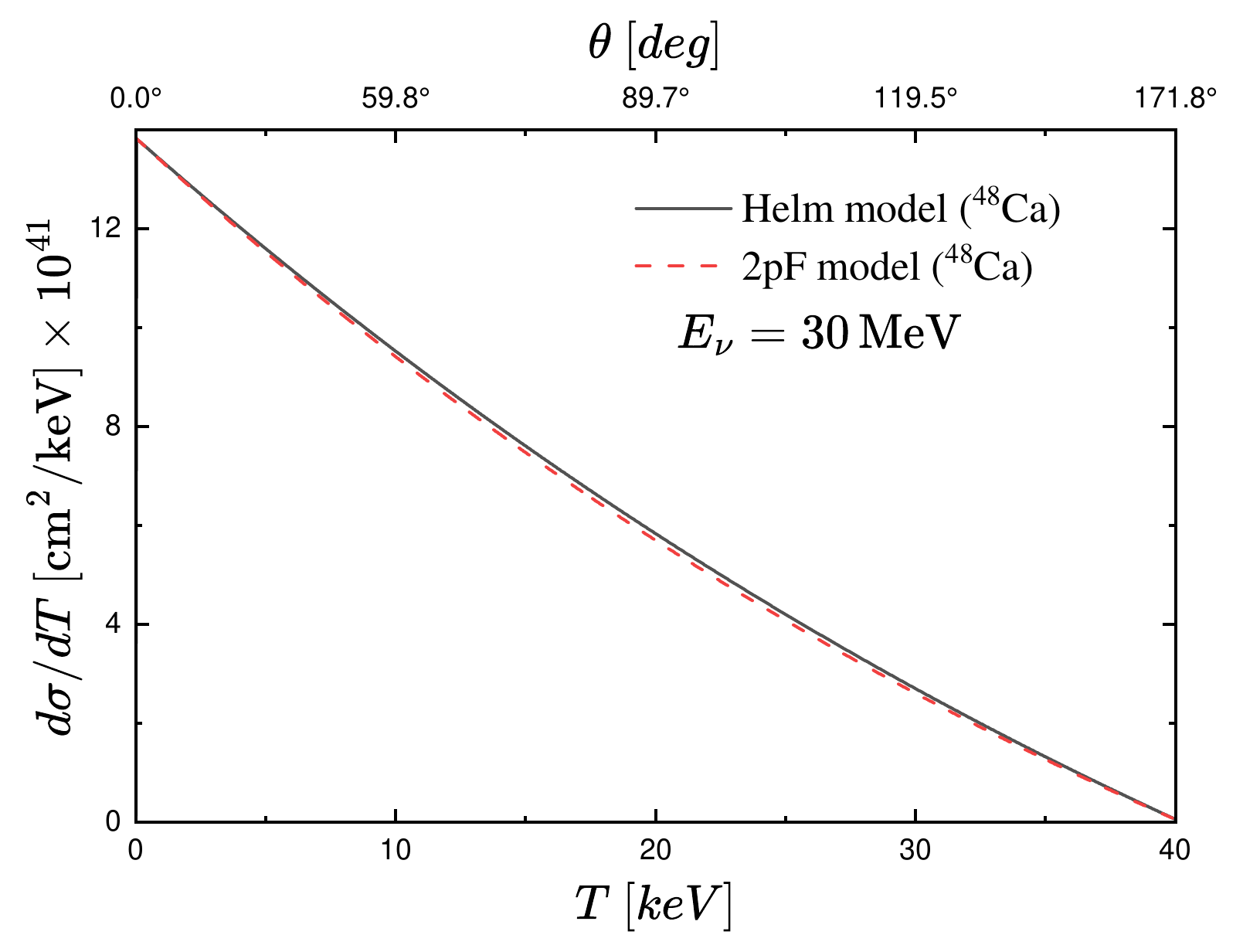}
	\includegraphics[width=0.49\linewidth]{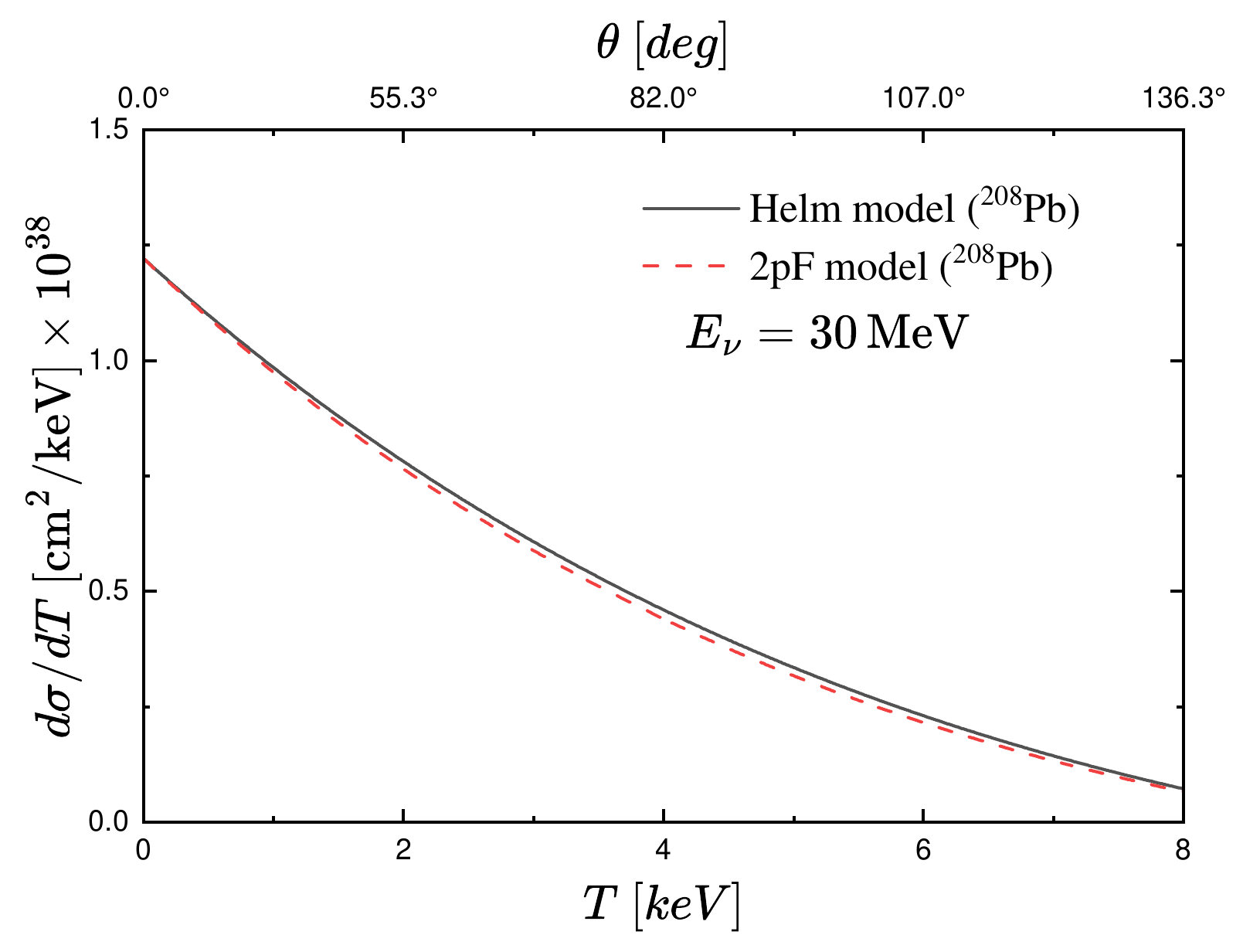}
\caption{
	Same as Fig.~\ref{fig:dsdT KDAR}, but for $\pi$DAR neutrinos
	($E_\nu\simeq30$~MeV).
    At these lower energies, the recoil spectra are dominated by the near--coherent region
	where $F(q)\simeq1$, and form--factor suppression (and Helm/2pF differences)
	becomes noticeable only in the high--recoil tail, especially for heavy targets.
}
\label{fig:dsdT piDAR}
\end{figure}

The comparison between KDAR and $\pi$DAR recoil spectra is shown in
Fig.~\ref{fig:dsdT piDAR}.
While both cases share the same low--$T$ rise driven by the weak--charge
normalization, the KDAR spectra extend to substantially higher recoil
energies and exhibit a progressively stronger suppression from
$|F(q)|^2$ as $qR$ approaches and exceeds unity.
This is precisely the kinematic region where sensitivity to nuclear
surface structure and neutron skins is maximal.
By contrast, $\pi$DAR spectra are dominated by the near--coherent regime, where nuclear--structure information is encoded only weakly.
Together with the nearly monoenergetic KDAR flux, which avoids the need
for spectral unfolding over a Michel distribution, these features make
KDAR uniquely well suited for precision CE$\nu$NS studies of neutron
radii and neutron skins.
\\

\noindent
Taken together, the results of this section demonstrate that KDAR--based
CE$\nu$NS accesses well--defined kinematic regions in which recoil spectra
exhibit genuine sensitivity to the shape of the nuclear weak form factor.
While these effects are modest at the level of individual spectra, they
arise in a controlled window beyond strict coherence.
In the following section, we make this connection quantitative by
developing a perturbative framework that relates variations of the weak
form factor to the neutron skin thickness.

\section{Perturbative interpretation of neutron-skin sensitivity in CE$\nu$NS}
\label{sec:skin_sensitivity}
The results presented in the previous section demonstrate that KDAR--based
CE$\nu$NS accesses kinematic regions in which recoil spectra exhibit genuine
sensitivity to the shape of the nuclear weak form factor.
In this section, we develop a quantitative and physically transparent
framework to interpret these spectral distortions in terms of the neutron
skin thickness,
$\Delta R_{np} \equiv R_n - R_p$,
where $R_n$ and $R_p$ denote the root--mean--square (rms) radii of the neutron
and proton distributions, respectively.

Rather than probing the neutron radius directly, CE$\nu$NS is sensitive to
the momentum--transfer dependence of the weak form factor $F_W(q)$.
Small variations of the neutron distribution modify $F_W(q)$ in a
systematic and momentum--dependent manner, particularly beyond the strictly
coherent regime.
The purpose of this section is therefore to formulate a perturbative
description that maps controlled variations of the weak form factor onto
variations of $\Delta R_{np}$, thereby clarifying how neutron--surface
information is encoded in CE$\nu$NS recoil spectra.

\subsection{Neutron dominance and a perturbative expansion of $F_W(q)$}
\label{subsec:fw_perturbative}
The weak form factor can be written as
\begin{equation}
	F_W(q)
	=
	\frac{
		g_p Z\,F_p(q) + g_n N\,F_n(q)
	}{
		g_p Z + g_n N
	},
\end{equation}
where $F_p(q)$ and $F_n(q)$ are the proton and neutron form factors, and
$g_p=\tfrac12-2\sin^2\theta_W$ and $g_n=-\tfrac12$ denote the vector weak couplings.
In the Standard Model, using $\sin^2\theta_W \simeq 0.2386$, the corresponding numerical
values are $g_p \simeq 0.023$ and $g_n = -0.5$, indicating that the proton weak coupling
is suppressed by more than an order of magnitude relative to the neutron one.
Throughout this work, we consistently adopt the low-energy value
$\sin^2\theta_W \simeq 0.2386$, appropriate for CE$\nu$NS kinematics.

It is therefore convenient to characterize the relative proton contribution to the weak
form factor by introducing the small, dimensionless expansion parameter
\begin{equation}
	\epsilon \equiv \frac{g_p Z}{g_n N},
	\label{eq:epsilon_def}
\end{equation}
which measures the ratio of proton to neutron contributions weighted by their weak
charges and particle numbers.
For the nuclei considered in this work, $\epsilon$ is negative and has a magnitude at the
level of only a few percent.
Numerically, this yields $|\epsilon|  \lesssim  O(0.05)$ for medium and heavy nuclei.
This confirms that the weak form factor probed in CE$\nu$NS is neutron dominated, while
proton contributions enter as a controlled perturbative correction.

Factoring out the dominant neutron contribution, the weak form factor can be rewritten as
\begin{equation}
	F_W(q)
	=
	\frac{F_n(q)+\epsilon F_p(q)}{1+\epsilon}.
\end{equation}
Expanding the denominator and retaining terms up to second order in $\epsilon$, one obtains
\begin{equation}
		F_W(q)
		\simeq
		F_n(q)
		-
		\epsilon\,[F_n(q)-F_p(q)]
		+
		\epsilon^2\,[F_n(q)-F_p(q)]
		\qquad
		(\mathcal{O}(\epsilon^3)\ \text{neglected}).
	\label{eq:fw_perturbative}
\end{equation}
This expression makes explicit that the weak form factor probed by CE$\nu$NS is neutron
dominated.
The leading correction arises at order $\epsilon$ and is proportional to the difference
between the proton and neutron form factors, while the $\epsilon^2$ term partially cancels
the linear contribution.
Given that $|\epsilon|$ is at the level of only a few percent for all nuclei considered
here, the expansion is rapidly convergent, and proton contributions enter as controlled
perturbative corrections.

\subsection{From weak form factor variations to neutron-skin variations}
\label{subsec:fw_to_skin}
The neutron skin thickness enters the weak form factor through the neutron radius,
which we parametrize as $R_n = R_p + \Delta R_{np}$.
In the following, derivatives with respect to $\Delta R_{np}$ are taken at fixed
proton rms radius $R_p$, so that variations of the neutron skin are attributed
entirely to changes in the neutron distribution.\footnote{
	In nuclear structure models, variations of $\Delta R_{np}$
	generally imply correlated changes of $R_p$ and $R_n$. Formally, the CE$\nu$NS
	response to $\Delta R_{np}$ involves a total derivative,
	\[
	\frac{d\mathcal{O}}{d(\Delta R_{np})}
	=
	\left(\frac{\partial \mathcal{O}}{\partial \Delta R_{np}}\right)_{R_p}
	+
	\left(\frac{\partial \mathcal{O}}{\partial R_p}\right)_{\Delta R_{np}}
	\frac{dR_p}{d(\Delta R_{np})},
	\]
	corresponding to correlated variations in $(R_p,R_n)$ space. The slope
	$dR_p/d(\Delta R_{np})$ and the associated covariance are model dependent and not
	experimentally constrained. We therefore restrict the present analysis to the
	conditional sensitivity at fixed $R_p$; correlated systematics and model
	comparisons are beyond the scope of this work (see, e.g., Refs.~\cite{Kim:2021skf, Mun:2024ked}).
}
This approximation is well justified: $R_p$ is experimentally constrained at the
level of $\sim\!0.01$--$0.02~\mathrm{fm}$ from elastic electron scattering, whereas
physically relevant variations of $\Delta R_{np}$ are typically
$\mathcal{O}(0.1~\mathrm{fm})$.
Possible correlated variations of $R_p$ are therefore subleading for CE$\nu$NS and
are neglected within the present sensitivity framework.

\subsubsection{Low-$q$ mapping and closed-form sensitivities}
\label{subsubsec:lowq_mapping}
We first establish an analytic benchmark for neutron-skin sensitivity in the
strictly coherent regime, $qR  \ll 1$, where the weak form factor admits a controlled
low-momentum expansion.
Using
\begin{equation}
	F_{p,n}(q)
	=
	1-\frac{q^2}{6}R_{p,n}^2
	+\frac{q^4}{120}\langle r^4\rangle_{p,n}
	+\mathcal{O}(q^6),
\end{equation}
and parametrizing the neutron radius as $R_n = R_p + \Delta R_{np}$, the weak form
factor expanded up to $\mathcal{O}(q^2)$ can be written as
\begin{equation}
	F_W(q)
	\simeq
	1-\frac{q^2}{6}
	\left[
	R_p^2
	+
	2R_p(1-\epsilon+\epsilon^2)\Delta R_{np}
	+
	(1-\epsilon+\epsilon^2)(\Delta R_{np})^2
	\right]
	+
	\mathcal{O}(q^4,\epsilon^3)
	+
	\mathcal{O}\!\left(q^2(\Delta R_{np})^3\right).
	\label{eq:FW_lowq_skin}
\end{equation}
Matching Eq.~(\ref{eq:FW_lowq_skin}) to the standard low-$q$ form
$F_W(q)\simeq 1-q^2R_W^2/6$ defines the weak-charge radius as
\begin{equation}
	R_W^2
	=
	R_p^2
	+
	2R_p\,a\,\Delta R_{np}
	+
	a\,(\Delta R_{np})^2
	+
	\mathcal{O}\!\left((\Delta R_{np})^3,\epsilon^3\right),
	\qquad
	a\equiv(1-\epsilon+\epsilon^2).
	\label{eq:RW2_lowq}
\end{equation}

For a reference value $\Delta R_{np}^{(0)}$ and a small shift
$\Delta R_{np}=\Delta R_{np}^{(0)}+\delta$, the recoil-spectrum distortion is defined as
\begin{equation}
	\Delta_\sigma(T;\delta)
	\equiv
	\frac{
		\left.\dfrac{d\sigma}{dT}\right|_{\Delta R_{np}^{(0)}+\delta}
	}{
		\left.\dfrac{d\sigma}{dT}\right|_{\Delta R_{np}^{(0)}}
	}
	-1,
	\qquad
    \ q = \frac{\sqrt{2MT}}{\hbar c}.
	\label{eq:delta_sigma_def}
\end{equation}
The logarithmic sensitivity coefficients,
\begin{equation}
	S_1(T)
	\equiv
	2\left.
	\frac{\partial}{\partial(\Delta R_{np})}\ln F_W(q)
	\right|_{\Delta R_{np}^{(0)}},
	\qquad
	S_2(T)
	\equiv
	2\left.
	\frac{\partial^2}{\partial(\Delta R_{np})^2}\ln F_W(q)
	\right|_{\Delta R_{np}^{(0)}},
\end{equation}
can then be evaluated analytically using Eq.~(\ref{eq:FW_lowq_skin}).
Keeping terms consistently up to $\mathcal{O}(q^2)$ yields
\begin{equation}
	S_1(T)
	\simeq
	-\frac{q^2}{3}\,a\left(R_p+\Delta R_{np}^{(0)}\right)
	+\mathcal{O}(q^4),
	\label{eq:S1_lowq}
\end{equation}
and
\begin{equation}
	S_2(T)
	\simeq
	-\frac{2q^2}{3}\,a
	-\frac{2q^4}{9}a^2\left(R_p+\Delta R_{np}^{(0)}\right)^2
	+\mathcal{O}(q^6),
	\label{eq:S2_lowq}
\end{equation}
where the $\mathcal{O}(q^4)$ term in $S_2$ originates from the
$(F_W'/F_W)^2$ contribution in $\partial^2\ln F_W/\partial(\Delta R_{np})^2$.

Exponentiating the logarithmic expansion and retaining terms up to $\delta^2$,
the recoil-spectrum distortion in the coherent regime becomes
\begin{equation}
	\Delta_\sigma(T;\delta)
	=
	S_1(T)\,\delta
	+
	\frac{1}{2}\left[S_2(T)+S_1^2(T)\right]\delta^2
	+
	\mathcal{O}(\delta^3).
	\label{eq:delta_sigma_lowq}
\end{equation}
Substituting Eqs.~(\ref{eq:S1_lowq}) and (\ref{eq:S2_lowq}) and keeping terms up to
$\mathcal{O}(q^4\delta^2)$ yields the explicit closed form
\begin{equation}
	\Delta_\sigma(T;\delta)
	\simeq
	-\frac{q^2}{3}a\left(R_p+\Delta R_{np}^{(0)}\right)\delta
	-\frac{q^2}{3}a\,\delta^2
	-\frac{q^4}{18}a^2\left(R_p+\Delta R_{np}^{(0)}\right)^2\delta^2
	+
	\mathcal{O}(\delta^3,q^6\delta^2).
	\label{eq:delta_sigma_lowq_closed}
\end{equation}

Equations~(\ref{eq:S1_lowq})–(\ref{eq:delta_sigma_lowq_closed}) provide a fully analytic
description of neutron-skin sensitivity in the strictly coherent regime.
They will be used below as a benchmark to assess the breakdown of the low-$q$
approximation and to interpret the numerical results obtained with the full weak form
factor at $qR \sim 1$ and in the vicinity of the diffractive minima.

\subsubsection{Beyond the coherent limit: sensitivities from the full form factor}
\label{subsubsec:fullFF_mapping}
The low-$q$ expansion in the previous subsection is not applicable once the
momentum transfer enters the diffractive regime. We therefore evaluate the
neutron-skin response directly from the full weak form factor.

We use the recoil-spectrum distortion by the exact ratio
\begin{equation}
	1+\Delta_\sigma(T;\delta)
	=
	\frac{
		F_W^2\!\left(q;\Delta R_{np}^{(0)}+\delta\right)
	}{
		F_W^2\!\left(q;\Delta R_{np}^{(0)}\right)
	},
	\label{eq:delta_sigma_fullFF}
\end{equation}
where $\Delta R_{np}^{(0)}$ is the reference neutron-skin value and $\delta$ is a
small shift applied along the one-parameter $\Delta R_{np}$ deformation direction.
For each shifted value, $F_W(q;\Delta R_{np})$ is recomputed without introducing
any approximation in $q$.

The sensitivity coefficients $S_1(T)$ and $S_2(T)$ are defined as in
Eq.~(22), but are now evaluated numerically from the full form factor via
central differences,
\begin{align}
	S_1(T)
	&\simeq
	\frac{1}{\delta}
	\left[
	\ln F_W\!\left(q;\Delta R_{np}^{(0)}+\delta\right)
	-
	\ln F_W\!\left(q;\Delta R_{np}^{(0)}-\delta\right)
	\right],
	\nonumber\\
	S_2(T)
	&\simeq
	\frac{2}{\delta^2}
	\left[
	\ln F_W\!\left(q;\Delta R_{np}^{(0)}+\delta\right)
	-
	2\ln F_W\!\left(q;\Delta R_{np}^{(0)}\right)
	+
	\ln F_W\!\left(q;\Delta R_{np}^{(0)}-\delta\right)
	\right].
	\label{eq:S1S2_fullFF_fd}
\end{align}
When desired, $\Delta_\sigma(T;\delta)$ may be related to $S_1(T)$ and $S_2(T)$
through the same second-order expansion already given in Eq.~(25).
Near diffractive minima, $\Delta_\sigma(T;\delta)$ can be strongly enhanced because
the baseline form factor in Eq.~(\ref{eq:delta_sigma_fullFF}) is suppressed; however,
the absolute rate is simultaneously reduced by the same factor
$d\sigma/dT \propto F_W^2(q)$. Consequently, the experimental reach must be assessed
by combining the spectral distortion with event statistics, as shown in the
numerical results below.

\begin{figure}
	\centering
	\includegraphics[width=0.49\linewidth]{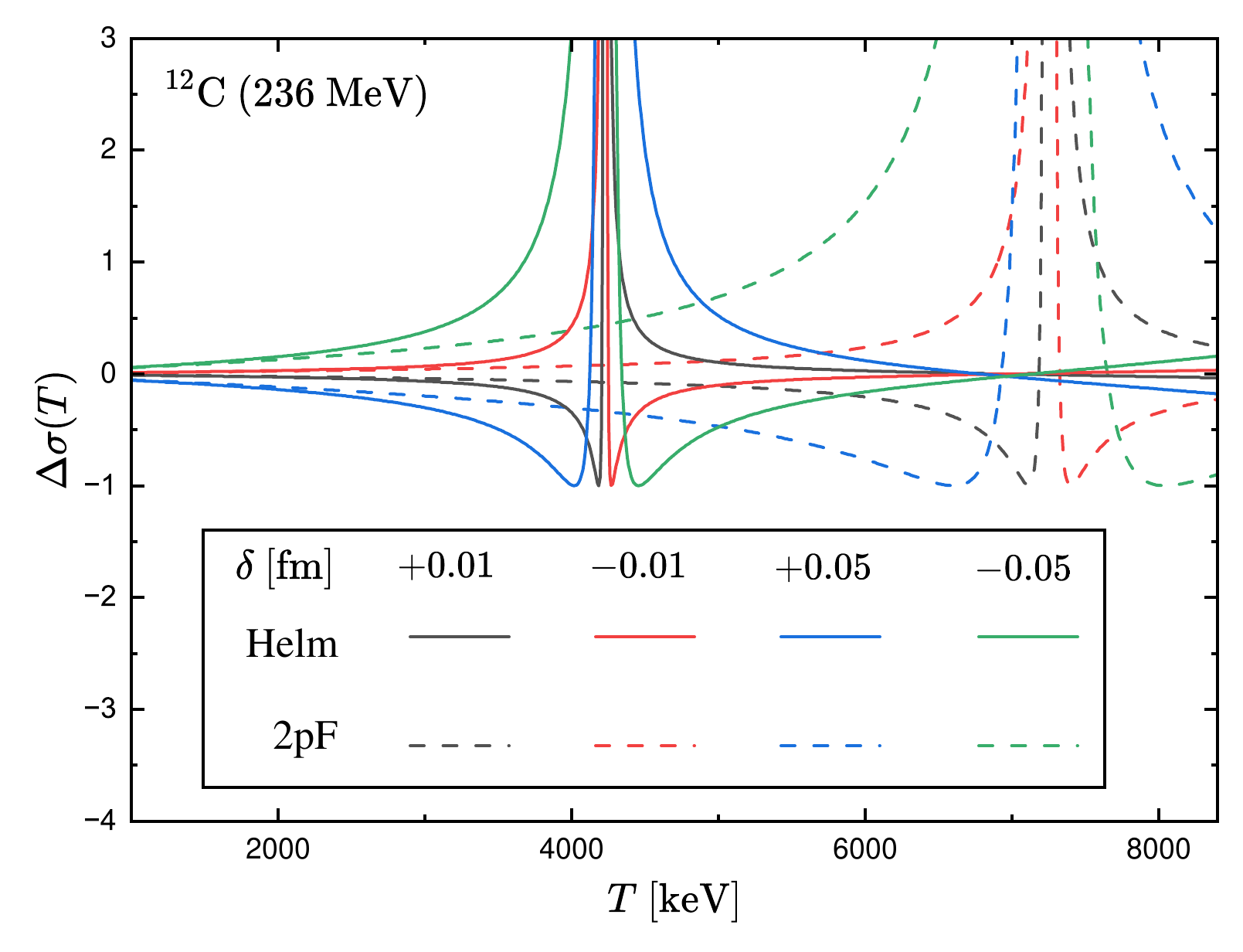}
    \includegraphics[width=0.49\linewidth]{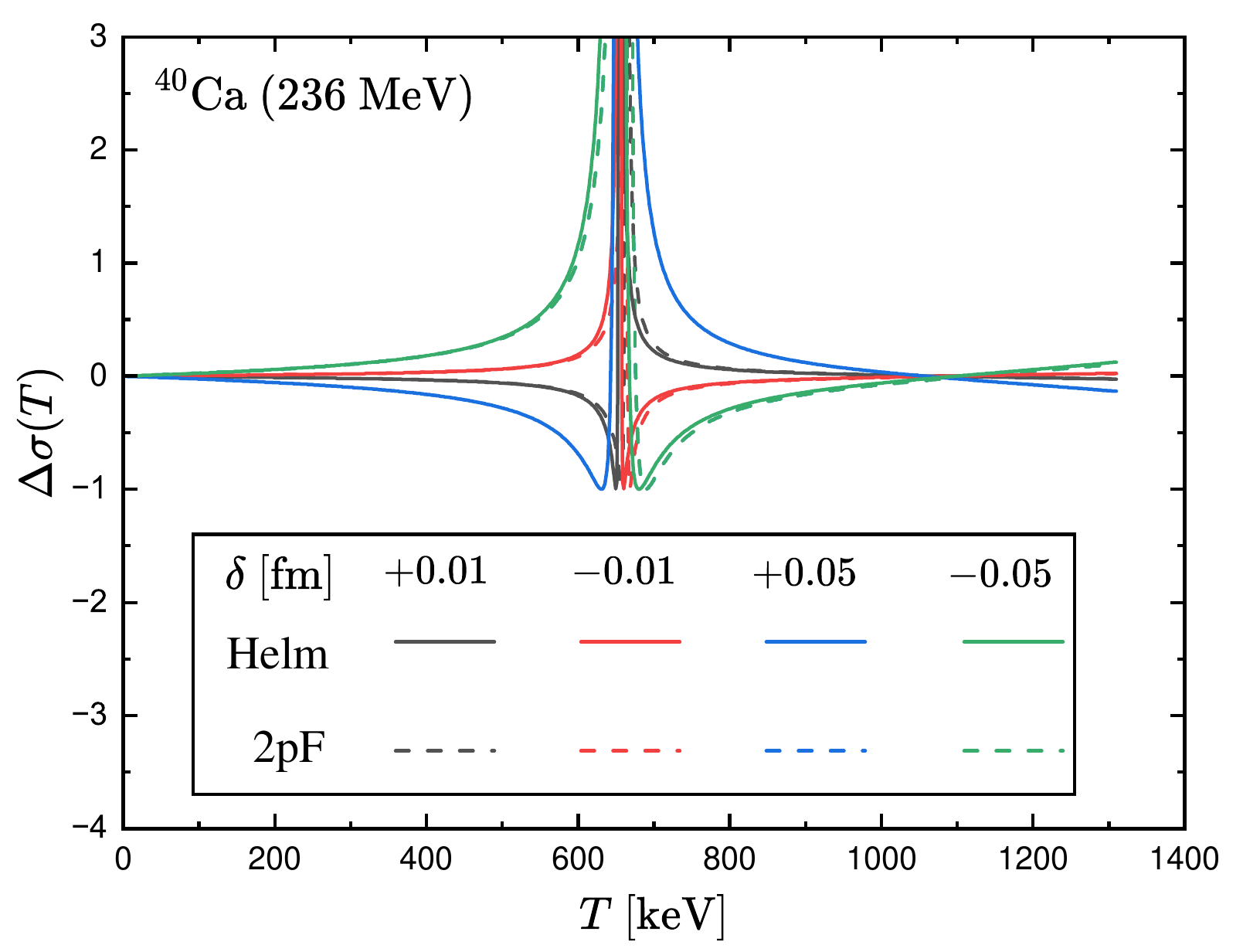}
    \includegraphics[width=0.49\linewidth]{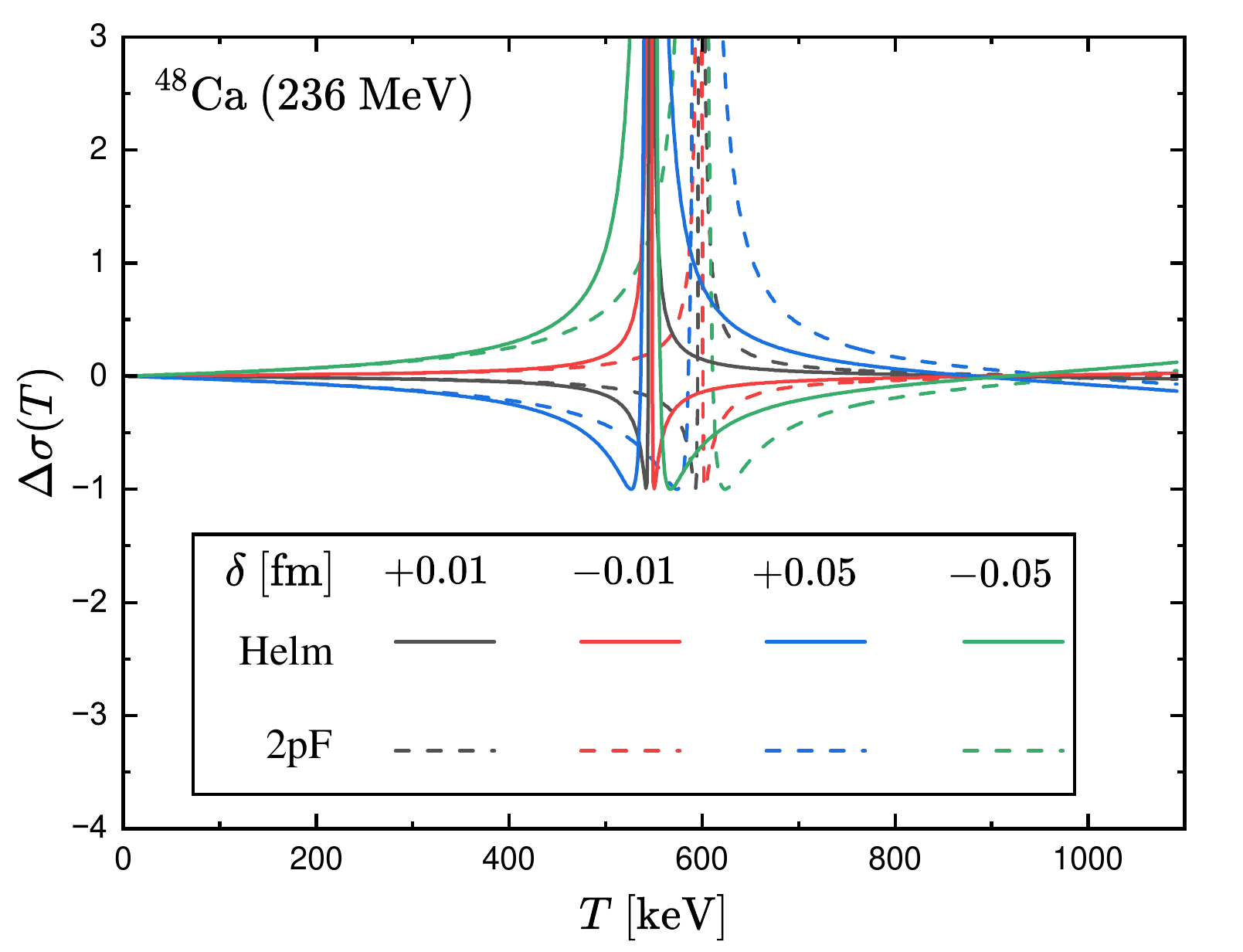}
    \includegraphics[width=0.49\linewidth]{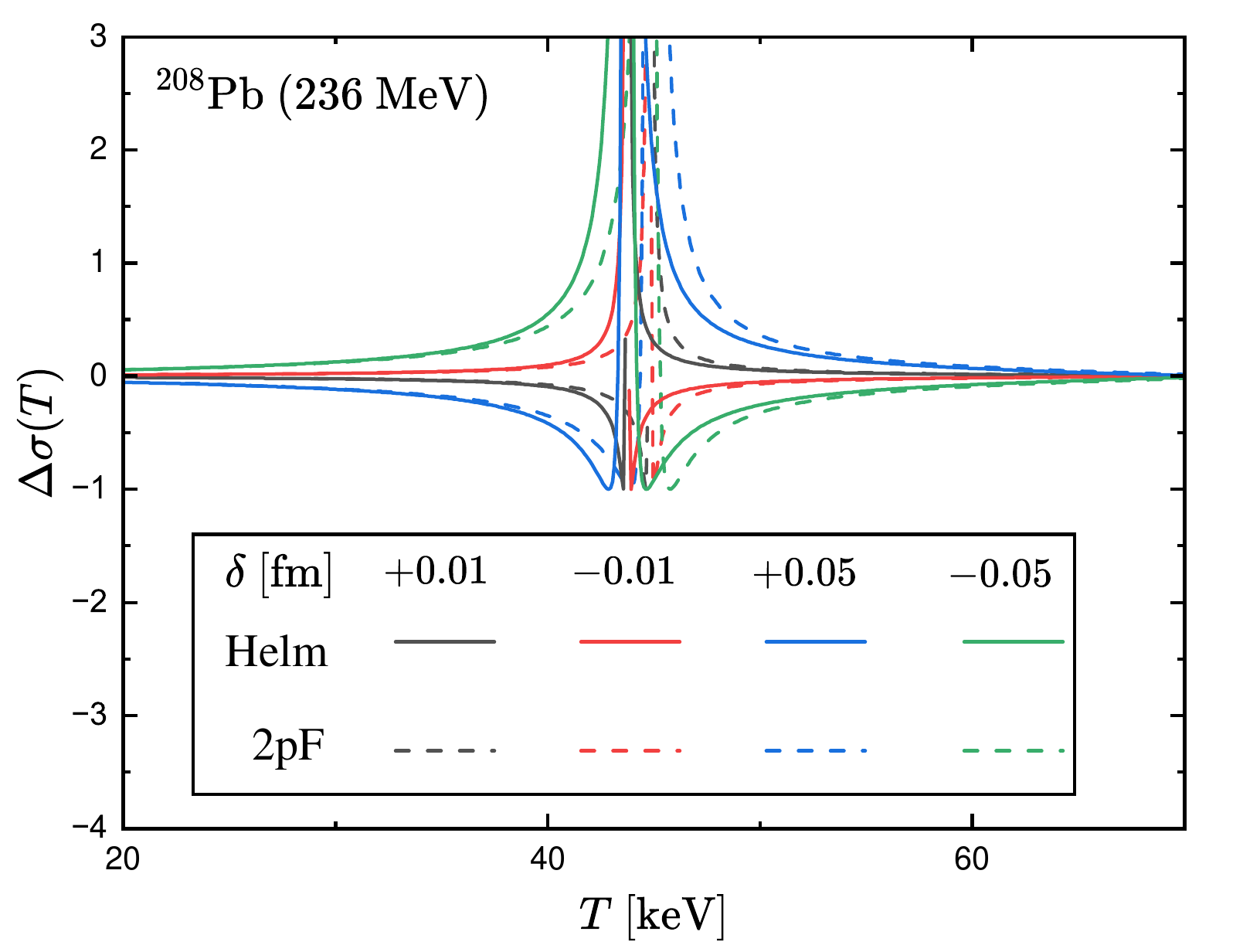}
\caption{
	Recoil-energy dependence of the fractional cross-section distortion
	$\Delta\sigma(T)$ induced by variations of the neutron-skin thickness
	$\delta=\Delta R_{np}-\Delta R_{np}^{(0)}$ for
	$^{12}\mathrm{C}$, $^{40}\mathrm{Ca}$, $^{48}\mathrm{Ca}$, and $^{208}\mathrm{Pb}$
	in CE$\nu$NS induced by KDAR $\nu_\mu$ with $E_\nu=236~\mathrm{MeV}$.
	Results for $\delta=\pm0.01~\mathrm{fm}$ and $\pm0.05~\mathrm{fm}$ are shown using
	the Helm (solid) and two-parameter Fermi (2pF, dashed) weak-charge form factors.
	At low recoil energies the distortion remains small and weakly model dependent,
	while near diffraction minima of the weak form factor it is strongly enhanced,
	reflecting the breakdown of the coherent limit.
	The recoil-energy locations of these enhancement regions depend on the nuclear size,
	leading to qualitatively different sensitivity patterns among light, medium, and heavy nuclei.
	This figure illustrates the kinematic localization and physical origin of
	neutron-skin sensitivity at KDAR energies rather than an optimized experimental observable.
}
	\label{fig:sensitivity}
\end{figure}

Figure~\ref{fig:sensitivity} shows the recoil--energy dependence of the fractional
spectral distortion $\Delta_\sigma(T;\delta)$ induced by shifting the neutron skin by
$\delta=\Delta R_{np}-\Delta R_{np}^{(0)}$, taking $\Delta R_{np}^{(0)}$ = 0~[fm]
as a reference.
The curves correspond to $\delta=\pm0.01~\mathrm{fm}$ and $\pm0.05~\mathrm{fm}$ and
compare the Helm (solid) and 2pF (dashed) parametrizations of the weak form factor.

In the low--recoil (near--coherent) region, $qR\lesssim1$, the distortion is small
and only weakly model dependent.
As $T$ increases and the kinematics enter the coherent--to--diffractive transition,
$F_W(q)$ departs from unity and the sensitivity grows; pronounced features can occur
near recoil energies that map onto diffractive structure of the form factor.
These large relative excursions do not necessarily imply a proportionally large experimental reach,
because the absolute rate is simultaneously suppressed by the same factor
$d\sigma/dT\propto F_W^2(q)$.
In contrast, calculations based on the full Helm form factor show pronounced
enhancements of $\Delta_\sigma$ at specific recoil energies.
These features are associated with diffractive minima of the weak form factor,
where $F_W(q)$ approaches zero and the relative cross--section change
\(
\Delta_\sigma(T;\delta)
=
\sigma(T;\Delta R_{np}^{(0)}+\delta)/
\sigma(T;\Delta R_{np}^{(0)})-1
\)
becomes strongly amplified.
The positions of these enhancements differ among nuclei, reflecting the
dependence of the diffractive pattern on the nuclear size and illustrating how
a fixed momentum--transfer condition maps onto different recoil energies for
different targets.

It is important to emphasize again that the large values of $\Delta_\sigma$ near
diffractive minima do not directly translate into proportionally large
experimental sensitivity.
In these regions the absolute cross section is strongly suppressed, since
$d\sigma/dT\propto F_W^2(q)$ is small, and the available event statistics are
limited.
The additional panel showing the fractional change obtained from the full form
factor alone makes this distinction explicit, separating regions of strong
relative sensitivity from those with appreciable event rates.

Overall, Fig.~\ref{fig:sensitivity} demonstrates that while the low--$q$
expansion reliably captures neutron--skin sensitivity in the strictly coherent
regime, it fails to describe the strong, structure--driven enhancements that
emerge once the kinematics enter the coherent--to--diffractive transition.
This comparison motivates the use of the full weak form factor in the following
section, where the recoil--spectrum sensitivity is quantified in terms of
experimentally meaningful observables and statistical reach.

\subsection{Physical constraints and quantitative sensitivity at KDAR}
\label{subsec:physical_and_quantitative_sensitivity}

\begin{table}
	\centering
	\small
	\caption{
		Momentum--transfer regimes probed by KDAR neutrinos for the nuclei
		considered in this work, together with representative experimental
		constraints on the neutron--skin thickness.
		The quoted $qR$ ranges indicate the region where the dominant
			contribution to the event rate arises; access to the first diffractive
		minimum at $qR\simeq4.493$ is kinematically possible at KDAR but occurs
		only in the high--recoil tail where the absolute rate is suppressed.
	}
	\label{tab:kdar_skin_summary}
	\resizebox{\textwidth}{!}{%
		\begin{tabular}{l c c c}
			\hline\hline
			Nucleus
			& Dominant $qR$ range
			& Sensitivity regime
			& Experimental constraints on $\Delta R_{np}$ (fm)
			\\
			\hline
			$^{12}$C
			& $\lesssim 1$
			& Near--coherent; form--factor distortions strongly suppressed
			& No direct neutron--skin constraint

			\\
			$^{40}$Ca
			& $\sim 1$
			& Coherent--transition region; onset of shape sensitivity
			& $-0.01^{+0.02}_{-0.02}$ (p--elastic scattering)
			\\
			$^{48}$Ca
			& $1$--$2$
			& Transition to weakly diffractive regime; enhanced sensitivity
			& $0.121\pm0.026~(\mathrm{exp})\pm0.024~(\mathrm{model})$ (CREX)
			\\
			$^{208}$Pb
			& $2$--$4$
			& Diffractive regime; strong form--factor modulation
			& $0.283\pm0.071$ (PREX--II)
			\\
			\hline\hline
		\end{tabular}
	}
\end{table}

We now present a quantitative estimate of the neutron-skin sensitivity achievable under realistic KDAR running conditions.
The relevant momentum-transfer regimes for each target nucleus have already been summarized
in Table~\ref{tab:kdar_skin_summary}; in this subsection we focus on the numerical sensitivity
results themselves, summarized in Table~\ref{tab:deltaR_sensitivity}.

The sensitivities are organized in terms of the time-integrated KDAR $\nu_\mu$ fluence at the
detector location, $\Phi_{\rm KDAR}$, defined as the total number of KDAR neutrinos crossing
unit area during the considered exposure.
For a stopped-kaon source, the fluence is obtained from the KDAR neutrino yield per proton on
target, the source--to--detector baseline, and the accumulated number of protons on target,
\begin{equation}
\Phi_{\rm KDAR}
=
\frac{Y_{\rm KDAR}}{4\pi L^2}\,N_{\rm POT}.
\label{eq:PhiKDAR_def}
\end{equation}
In the JSNS$^2$ benchmark adopted here, we use a baseline of $L=24$~m and a KDAR yield
$Y_{\rm KDAR}=3.8\times10^{-3}$~KDAR~$\nu_\mu$/POT, which corresponds to a fluence per POT of
$(\Phi_{\rm KDAR}/{\rm POT})=5.25\times10^{-11}~\mathrm{cm}^{-2}\mathrm{/POT}$.
Using the POT milestones reported by JSNS$^2$, 
this definition leads to the four benchmark
fluences employed in Table~\ref{tab:deltaR_sensitivity},
\begin{equation}
\Phi_{\rm KDAR}
=
6.82\times10^{11},\;
2.57\times10^{12},\;
5.85\times10^{12},\;
9.75\times10^{12}\ \mathrm{cm}^{-2},
\end{equation}
corresponding to the 2021 dataset~\cite{JSNS2:2024bzf}, 
the cumulative exposure as of mid--2024~\cite{JSNS2:2024bzf}, 
the approved three-year program~\cite{JSNS2:2020hmg}, 
and an extended five-year,\footnote{
	The five--year exposure is not quoted as an explicit POT value in the JSNS$^2$ documents.
	Instead, it is obtained by a linear extrapolation of the approved three--year exposure,
	$N_{\rm POT}=1.114\times10^{23}$ (corresponding to 1~MW $\times$ 3~years),
	assuming constant beam delivery.
	This yields $N_{\rm POT}^{(5\,{\rm yr})}=(5/3)\times1.114\times10^{23}
	=1.857\times10^{23}$.
	The use of a five--year scenario is motivated by the stated long--term operation goal
	of JSNS$^2$ at 1~MW beam power. \label{footnote:POT-5years}
} respectively.
\begin{table}
	\centering
	\caption{
		Statistical $1\sigma$ sensitivity to the neutron skin thickness $\Delta R_{np}$
		obtained from the condition $\Delta\chi^2=1$ using KDAR-based CE$\nu$NS.
		We consider representative integrated KDAR fluences at a JSNS$^2$-like facility
		corresponding to the 2021 dataset~\cite{JSNS2:2024bzf}, the cumulative exposure as of mid-2024~\cite{JSNS2:2024bzf},
		the approved three-year program~\cite{JSNS2:2020hmg}, 
        and an extended five-year exposure\,\textsuperscript{3}.
		Here $\Phi_{\rm KDAR}$ denotes the time-integrated KDAR $\nu_\mu$ fluence at the detector location,
		i.e., the total number of KDAR neutrinos crossing unit area during the considered exposure,
		quoted in units of cm$^{-2}$.
		Unless stated otherwise, the estimates assume a recoil threshold of
		$T_{\rm th}=10$~keV, Gaussian energy smearing with
		$\sigma_T/T = 10\%/\sqrt{T/\mathrm{MeV}}$, and a total exposure of
		10~ton$\cdot$year.
	}
	\label{tab:deltaR_sensitivity}
\begin{tabular}{lcccc}
	\hline
	& \multicolumn{4}{c}{$\Delta R_{np}^{\,(1\sigma)}$ [fm]} \\
	\cline{2-5}
	Exposure
	& 2021 & mid-2024 & 3-year & 5-year \\
	Target nucleus $\backslash$ $\Phi_{\rm KDAR}\ [{\rm cm}^{-2}]$
	& $6.82\times10^{11}$ & $2.57\times10^{12}$ & $5.85\times10^{12}$ & $9.75\times10^{12}$ \\
	\hline
	$^{12}$C
	& $0.13048$ & $0.06698$ & $0.04420$ & $0.03409$ \\
	$^{40}$Ca
	& $0.12063$ & $0.06201$ & $0.04106$ & $0.03167$ \\
	$^{48}$Ca
	& $0.09257$ & $0.04749$ & $0.03141$ & $0.02398$ \\
	$^{208}$Pb
	& $0.06835$ & $0.03489$ & $0.02291$ & $0.01758$ \\
	\hline
\end{tabular}
\end{table}

Table~\ref{tab:deltaR_sensitivity} reports the projected statistical $1\sigma$ precision
$\Delta R_{np}^{\,(1\sigma)}$ obtained from a binned $\chi^2$ analysis of KDAR-induced recoil spectra,
defined by the condition $\Delta\chi^2=1$.
Unless stated otherwise, the calculation assumes a recoil threshold of $T_{\rm th}=10$~keV,
Gaussian energy smearing $\sigma_T/T=10\%/\sqrt{T/\mathrm{MeV}}$, and a total exposure of
10~ton$\cdot$year; the numerical procedure is detailed in
Appendix~\ref{app:numerical_sensitivity}.
The numerical values in Table~\ref{tab:deltaR_sensitivity} provide a fully quantitative
statement of the neutron-skin reach attainable with KDAR-based CE$\nu$NS.
For the medium-mass nucleus $^{48}$Ca, the projected precision improves from
$\Delta R_{np}^{\,(1\sigma)}=0.0926~\mathrm{fm}$ at the 2021 benchmark fluence to
$0.0240~\mathrm{fm}$ in the extended five-year scenario.
This nearly fourfold improvement tracks the increase in $\Phi_{\rm KDAR}$ and confirms that
the extraction remains statistics dominated over the considered exposure range.
A similar trend is observed for $^{40}$Ca, whose sensitivity closely follows that of
$^{48}$Ca at all benchmark fluences.

For the heavy nucleus $^{208}$Pb, Table~\ref{tab:deltaR_sensitivity} shows that precisions at
the level of $0.02~\mathrm{fm}$ become attainable as the KDAR fluence accumulates
(e.g.\ $0.0684~\mathrm{fm}$ at the 2021 benchmark improving to $0.0176~\mathrm{fm}$ in the
five-year scenario).
Although the relative recoil-spectrum distortions induced by neutron-skin variations are
enhanced in the diffractive regime for heavy targets, the information is concentrated in the
high-recoil tail where the absolute event rate is suppressed.
As a consequence, the final statistical precision for $^{208}$Pb becomes comparable to,
rather than parametrically stronger than, that obtained for the optimal medium-mass target.
By contrast, the light nucleus $^{12}$C remains intrinsically less constraining.
Even at the largest benchmark fluence, the projected uncertainty,
$\Delta R_{np}^{\,(1\sigma)}=0.0341~\mathrm{fm}$, is significantly weaker than for the heavier
targets, reflecting the dominance of near-coherent kinematics where neutron-skin effects
enter primarily through an overall normalization.

Taken together, Table~\ref{tab:deltaR_sensitivity} defines the quantitative experimental
parameter space for KDAR-based neutron-skin measurements at a JSNS$^2$-like facility.
It demonstrates, in explicitly numerical terms, how the achievable precision on
$\Delta R_{np}$ scales with accumulated KDAR fluence and identifies medium-mass nuclei,
in particular $^{48}$Ca, as providing the optimal balance between event statistics and
shape sensitivity under realistic detector assumptions.

\section{Summary and Conclusions}
In this work, we have developed a unified and quantitatively controlled
framework that connects nuclear weak--charge densities, weak form factors,
and observable CE$\nu$NS recoil spectra across the transition from strict
coherence to nuclear diffraction.
Focusing on KDAR neutrinos with
$E_\nu = 236~\mathrm{MeV}$, we have systematically contrasted their kinematic
reach with that of conventional pion--decay--at--rest ($\pi$DAR) neutrinos
and demonstrated the unique advantages of KDAR for probing neutron--skin
effects in nuclei.
Our analysis spans light, medium--mass, and heavy targets,
$^{12}$C, $^{40}$Ca, $^{48}$Ca, and $^{208}$Pb, allowing for a coherent
comparison across nuclear size.

A central result of this study is that CE$\nu$NS sensitivity to neutron--skin
thickness is strongly governed by the dimensionless variable $qR$.
In the strictly coherent regime ($qR \lesssim 1$), nuclear--structure effects
are suppressed and neutron--skin variations manifest primarily as overall
normalization changes.
Meaningful sensitivity emerges only once the kinematics extend into the
coherent--to--diffractive transition, where the weak form factor departs from
unity and induces genuine, shape--dependent distortions in the recoil
spectrum.
The vicinity of the first diffractive minimum at $qR \simeq 4.493$ therefore
provides a natural benchmark for maximal sensitivity to neutron surface
properties.

Using both the Helm and two-parameter Fermi (2pF) parametrizations normalized to the same
weak charge, we find that they reproduce comparable overall size scales and surface locations
for the nuclei considered, while differing in how the weak charge is distributed between the
core plateau and the diffuse surface.
Accordingly, the form factors are nearly identical in the deep-coherence limit, but begin to
separate once the kinematics enter the coherence--transition region, where recoil spectra
acquire genuine shape dependence on the nuclear surface.

These differences remain negligible in the low--$q$ limit yet translate into
systematic and physically interpretable spectral distortions once
$qR \gtrsim 1$.
This observation underscores the necessity of employing the full weak form
factor, rather than low--$q$ expansions, when interpreting KDAR--induced
CE$\nu$NS spectra.

Exact elastic kinematics further reveal a qualitative contrast between KDAR
and $\pi$DAR measurements.
While $\pi$DAR--based CE$\nu$NS remains confined to the near--coherent regime
for all nuclei considered, KDAR neutrinos extend the accessible
momentum--transfer range into $qR \gtrsim 1$ for light and medium--mass nuclei
and provide access to the diffractive regime for heavier targets.
This difference explains why KDAR--based CE$\nu$NS uniquely enables
shape--sensitive probes of nuclear structure beyond normalization--only
observables.

Building on these physical insights, we have presented a quantitative
estimate of the statistical sensitivity to the neutron skin thickness
achievable at a JSNS$^2$--like facility.
For a total exposure of 10~ton$\cdot$year and representative integrated KDAR
fluences, the projected $1\sigma$ sensitivities reach
$\Delta R_{np}^{\,(1\sigma)}\simeq0.09$--$0.02$~fm for $^{48}$Ca as the fluence
increases, while sensitivities for $^{208}$Pb approach
$\Delta R_{np}^{\,(1\sigma)}\simeq0.07$--$0.02$~fm.
In contrast, lighter nuclei such as $^{12}$C remain limited by their
proximity to the coherent regime.
These results demonstrate that KDAR CE$\nu$NS can probe neutron--skin effects
through genuine spectral distortions rather than overall rate changes.

In summary, KDAR--based CE$\nu$NS transforms elastic
neutrino--nucleus scattering from a predominantly coherent probe governed by
weak--charge scaling into a diffraction--sensitive measurement of the nuclear
weak form factor.
The nearly monoenergetic nature of the KDAR source provides a clean and
transparent mapping between momentum transfer, nuclear size, and recoil
spectrum distortions, enabling controlled access to neutron surface
information.
When combined with existing parity--violating electron--scattering results
and multi--target analyses, KDAR CE$\nu$NS offers a complementary and
electroweakly clean avenue for precision studies of neutron radii and
neutron--skin thicknesses in nuclei.

\section*{Acknowledgments}
This work was supported by the National Research Foundation of Korea
(Grant Nos. 
RS-2022-NR069287, RS-2022-NR070836, 
RS-2024-00460031, 
RS-2025-00513410, RS-2025-24533596, and RS-2025-25400847).
The work of MKC was supported by the National Research Foundation of Korea
(Grant Nos. RS-2021-NR060129 and RS-2025-16071941).

\appendix

\section{Elastic kinematics and the variable $qR$}
\label{app:kinematics}
We summarize the kinematic relations used in the main text for the
neutral-current elastic process
$\nu(k)+A(P)\to \nu(k')+A(P')$.
We work in the laboratory frame with the target at rest,
$P^\mu=(M,\mathbf{0})$, and metric $g^{\mu\nu}=\mathrm{diag}(+,-,-,-)$.
The nuclear recoil kinetic energy is
$T \equiv P^{\prime 0}-M$.

\subsection{Momentum transfer and recoil energy}
In this Appendix, $q^\mu$ denotes the four-momentum transfer, whereas
$q$ denotes the corresponding wavenumber in fm$^{-1}$ defined in Eq.~(\ref{eq:app_q_fm}).
Define the four-momentum transfer
$q^\mu \equiv k^\mu-k^{\prime\mu}=P^{\prime\mu}-P^\mu$ and
$Q^2\equiv -q^2>0$.
Using $P'^2=P^2=M^2$, one finds the exact elastic identity
	\begin{equation}
		m_\ell^2 + 2\,E_\nu\,M \;-\; 2(E_\nu+M)\,E_\nu' \;+\; 2E_\nu\,|\mathbf{k}'|\cos\theta \;=\; 0,
		\label{eq:master}
	\end{equation}
	Here $\theta$ denotes the scattering angle of the outgoing neutrino,
	$\cos\theta \equiv \hat{\bm{k}}\cdot\hat{\bm{k}}'$, 
	and the exact four–momentum transfer
	
	\begin{equation}
		|\mathbf{q}|^2 \;\equiv\; |(k-k')|^2 \;=\; - m_\ell^2 + 2E_\nu E_\nu' - 2 E_\nu k' \cos\theta 
		\;\equiv\; Q^2 .
		\label{eq:Q2exact}
	\end{equation}.
	
	When for a (nearly) massless neutrino ($m_\ell\to0$), Eq.~\eqref{eq:Q2exact} reduces to, 
	\begin{equation}
		Q^2 \simeq 2E_\nu E_\nu'(1-\cos\theta), 
		\label{eq:app_Q2_angle}
	\end{equation}	
	Eq.~\eqref{eq:master} reduces to 
	\begin{equation}
		2E_\nu E_\nu'(1-\cos\theta) \simeq 2MT.
		\label{eq:app_2MT_angle}
	\end{equation}
	
	So in the heavy-target CEvNS limit (nonrelativistic recoil),

\begin{equation}
	Q^2 \simeq 2MT .
	\label{eq:app_Q2}
\end{equation}
For CE$\nu$NS, we may use the standard approximation
$|\mathbf{q}|\simeq \sqrt{2MT}$ when evaluating the weak form factor.

For convenience, the momentum transfer in \si{fm^{-1}} is
\begin{equation}
	q~[\si{fm^{-1}}] \equiv |\mathbf{q}|
	\simeq \sqrt{2MT}.
	\label{eq:app_q_fm}
\end{equation}

\subsection{Angle--recoil mapping and kinematic endpoint}
Combining $E_\nu' = E_\nu - T$ with Eq.~\eqref{eq:app_Q2}  
and Eq.~\eqref{eq:app_Q2_angle} yields 
\begin{equation}
	\cos\theta(T)
	= 1 - \frac{MT}{E_\nu(E_\nu-T)} .
	\label{eq:app_costheta_T}
\end{equation}
Equivalently,
\begin{equation}
	T(E_\nu,\theta)
	= \frac{E_\nu^2(1-\cos\theta)}{M+E_\nu(1-\cos\theta)} .
	\label{eq:app_T_theta}
\end{equation}
The maximum recoil occurs for backward scattering ($\theta=\pi$),
\begin{equation}
	T_{\max}(E_\nu,M)=\frac{2E_\nu^2}{M+2E_\nu} .
	\label{eq:app_Tmax}
\end{equation}
Inverting Eq.~\eqref{eq:app_Tmax}, the minimum neutrino energy required to
produce a recoil $T$ is
\begin{equation}
	E_{\nu,\min}(T)
	= \frac12\left(T+\sqrt{T^2+2MT}\right)~,
	\label{eq:app_Enu_min}
\end{equation}
 and  the final lepton energy $E_\nu'$ is
	\begin{equation}
		E_\nu'(T,\theta)
		= \frac{E_\nu}{1+\frac{E_\nu}{M}(1-\cos\theta)}.
		\label{eq:final_lepton_E}
	\end{equation}
	with Eq.~\eqref{eq:app_T_theta} and $E_\nu' = E_\nu - T$ .

\subsection{Dimensionless coherence variable $qR$ and benchmark scales}
The degree of coherence is controlled by the dimensionless combination
\begin{equation}
	qR \equiv |\mathbf{q}|\,R
	\simeq \sqrt{2MT}\,R,
	\label{eq:app_qR}
\end{equation}
where $R$ is a characteristic nuclear radius (e.g.\ $R=r_0A^{1/3}$ with
$r_0\simeq 1.2~\si{fm}$, as used for Table~\ref{tab:scales}).
This $R$ is introduced only to form the dimensionless variable $qR$;
the weak form factor itself is computed with its own model parameters
(Helm: $R_0,s$; 2pF: $c,a$).

For quick estimates, Eq.~\eqref{eq:app_qR} gives the characteristic recoil
scale at a chosen $qR=x$,
\begin{equation}
	T(qR=x)\simeq \frac{x^2}{2MR^2}\ .
	\label{eq:app_T_of_qR}
\end{equation}
Near the kinematic endpoint (backward scattering) and for $E_\nu\ll M$,
the neutrino energy that reaches $qR=x$ is approximately
\begin{equation}
	E_\nu(qR=x)\simeq \frac{x}{2R},
	\qquad (x=1,\ 4.493,\ldots).
	\label{eq:app_Enu_of_qR}
\end{equation}
The kinematic endpoint is set by $T_{\max}(E_\nu,M)$ in Eq.~(\ref{eq:app_Tmax}).
Separately, diffractive minima arise from the form factor: for a sharp-sphere benchmark
the first minimum occurs at $qR\simeq 4.493$ (and the next at $qR\simeq 7.725$),
providing a convenient reference scale for maximal form-factor discrimination.

\section{Absolute cross-section variations induced by neutron-skin changes}

\begin{figure}
	\centering
	\includegraphics[width=0.49\linewidth]{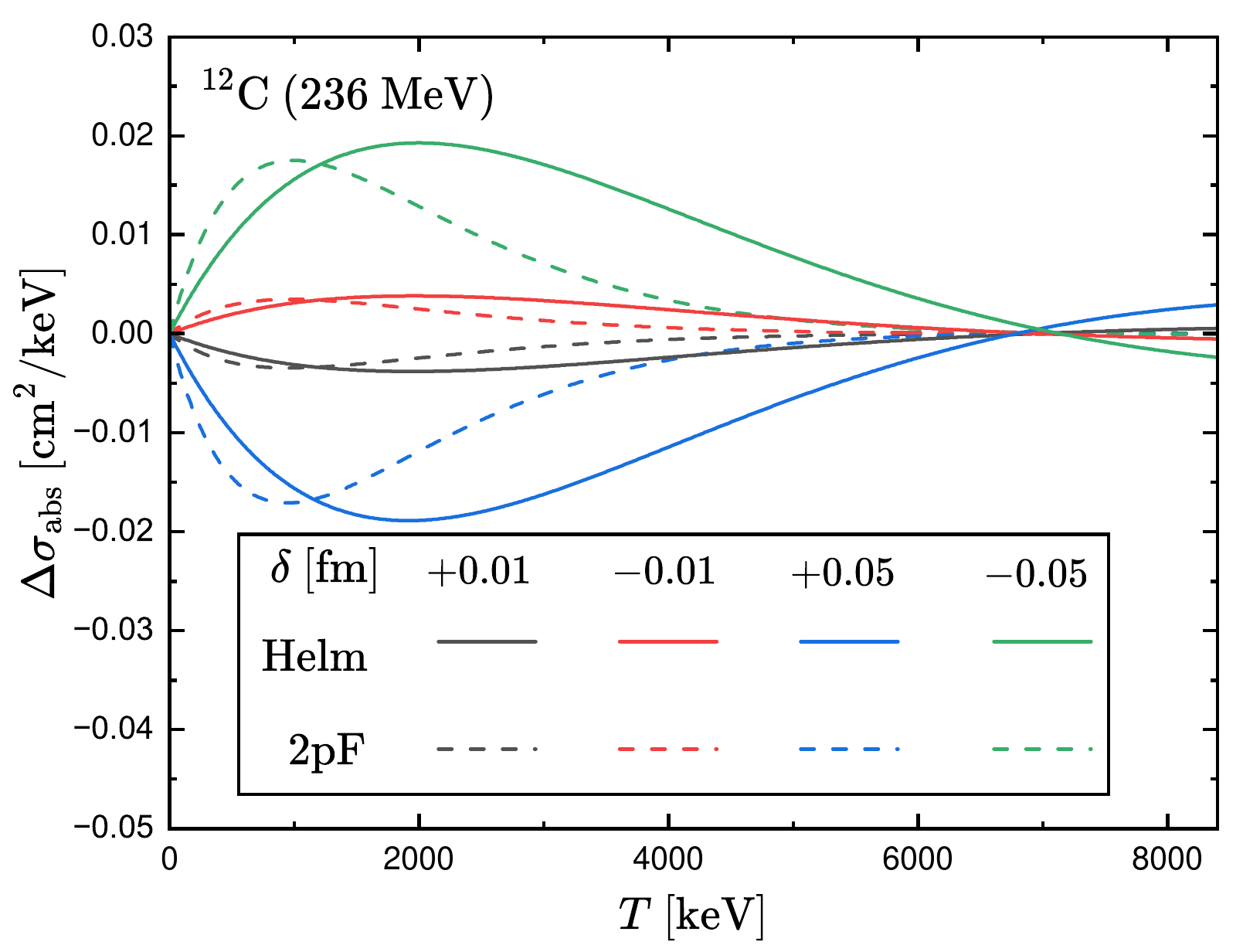}
    \includegraphics[width=0.49\linewidth]{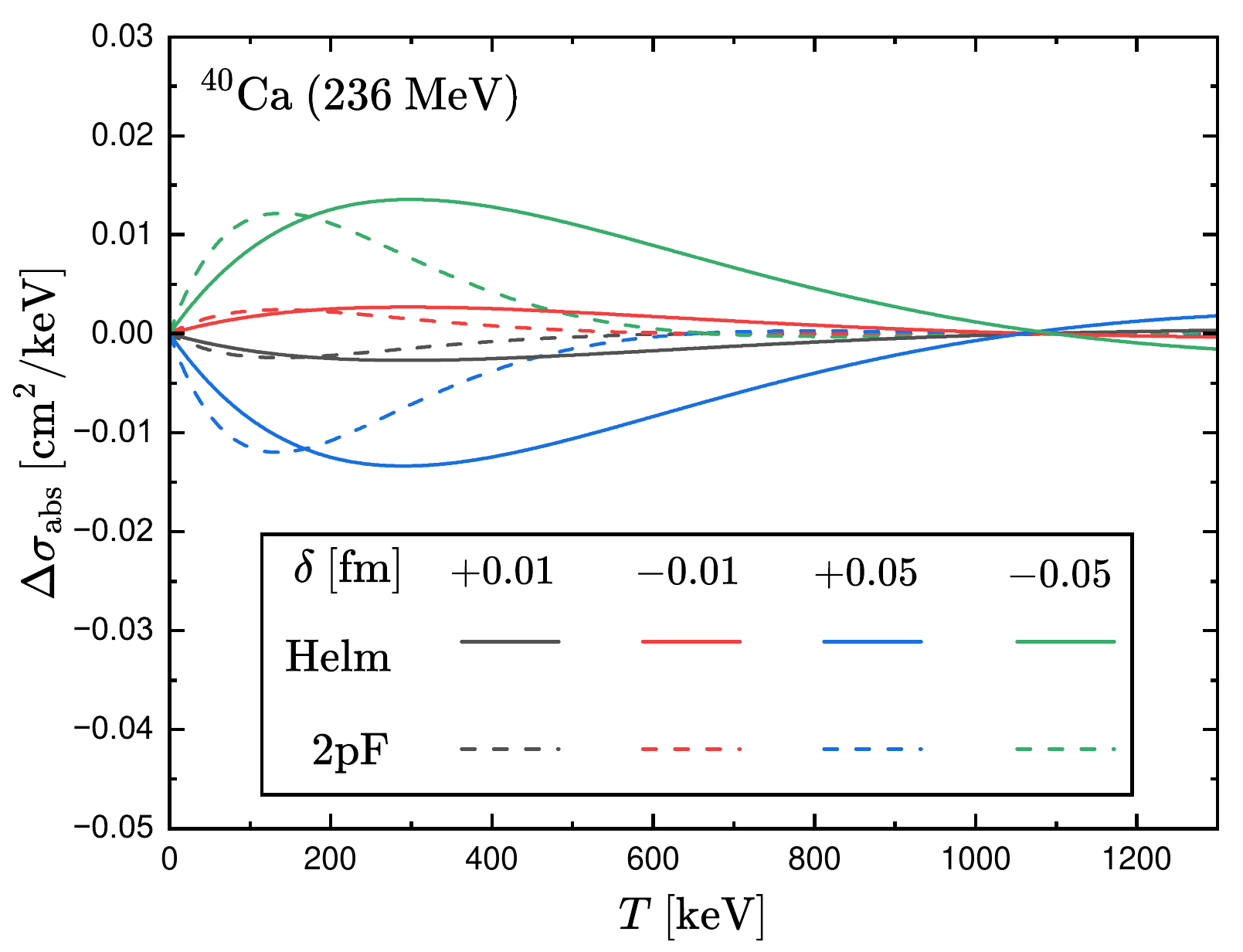}
    \includegraphics[width=0.49\linewidth]{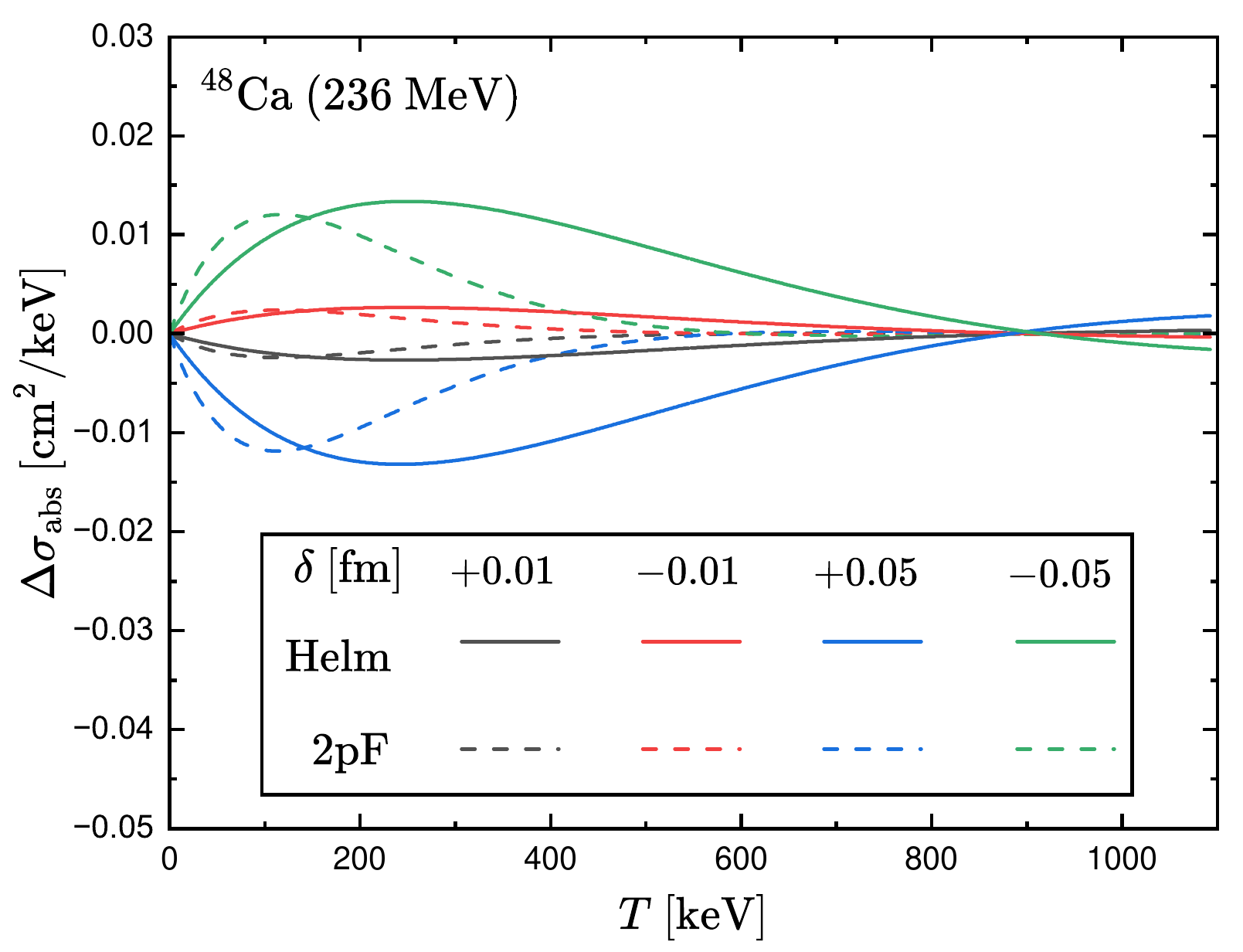}
    \includegraphics[width=0.49\linewidth]{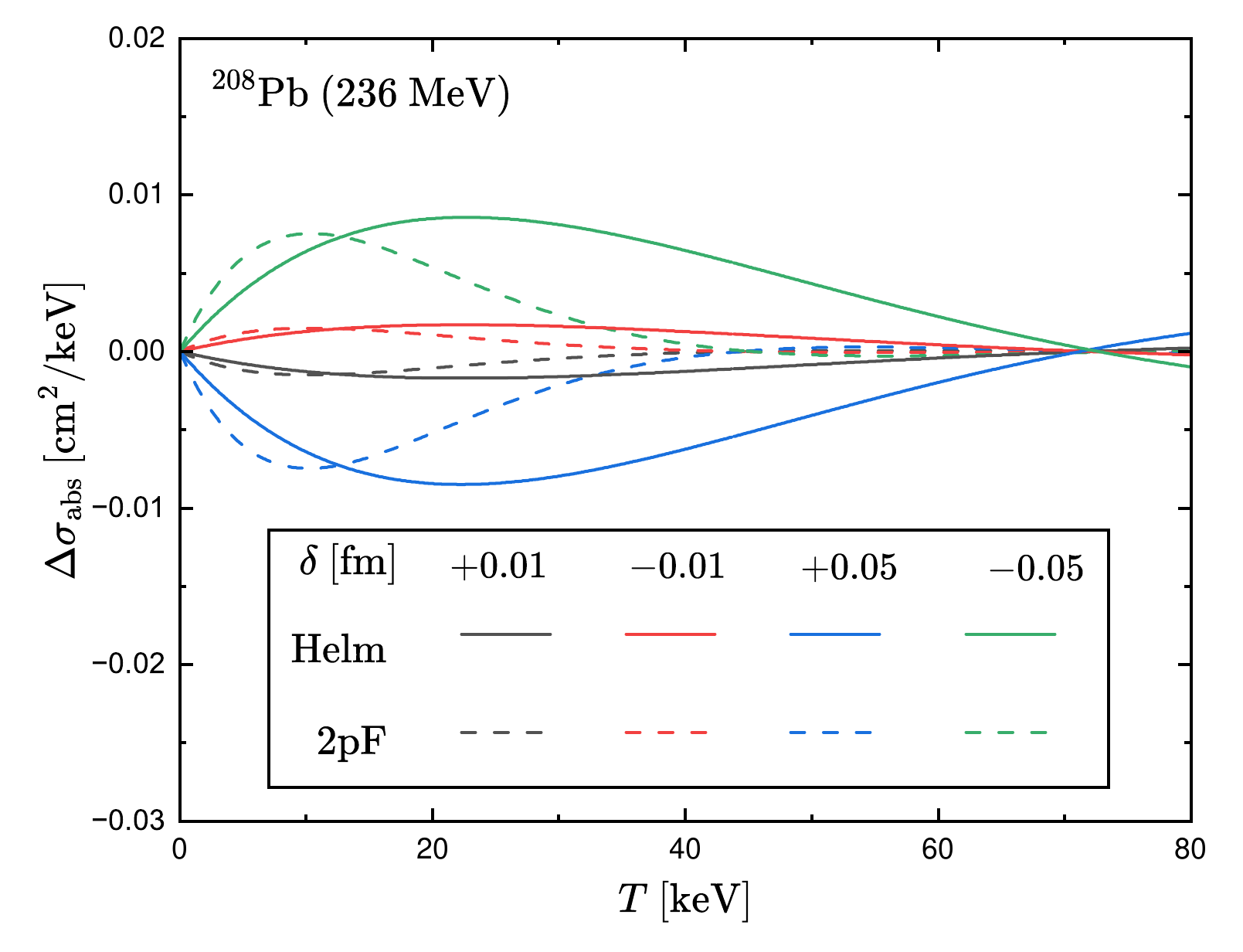}
\caption{
	Absolute difference in the CE$\nu$NS differential cross section,
	$\Delta_{\rm abs}(T)=d\sigma/dT(T;\Delta R_{np}^{(0)}+\delta)
	- d\sigma/dT(T;\Delta R_{np}^{(0)})$,
	for KDAR $\nu_\mu$ with $E_\nu=236~\mathrm{MeV}$.
	Results are shown for $^{12}$C, $^{40}$Ca, $^{48}$Ca, and $^{208}$Pb using
	the Helm (solid) and two-parameter Fermi (2pF, dashed) weak-charge form factors.
	The absolute scale illustrates the magnitude of the recoil-spectrum variation,
	complementing the fractional distortion discussed in the main text.
}
	\label{fig:sensitivity_diff}
\end{figure}

\label{app:absolute_difference}
In the main text, the sensitivity to neutron-skin variations is illustrated
in terms of the fractional distortion of the recoil spectrum,
which highlights the intrinsic dependence on the nuclear weak form factor.
To complement this representation, Fig.~\ref{fig:sensitivity_diff}
shows the corresponding absolute difference in the differential cross section,
\begin{equation}
	\Delta \sigma_{\rm abs}(T)
	=
	\frac{d\sigma}{dT}(T;\Delta R_{np}^{(0)}+\delta)
	-
	\frac{d\sigma}{dT}(T;\Delta R_{np}^{(0)}),
\end{equation}
for $\delta=\pm0.01~\mathrm{fm}$ and $\pm0.05~\mathrm{fm}$.

As expected, the absolute difference remains finite and smooth across the
entire recoil-energy range, including the vicinity of diffractive minima
where the fractional distortion exhibits pronounced enhancements.
This behavior reflects the fact that, although the relative sensitivity is
maximized near form-factor nodes, the cross section itself is suppressed
in these regions.
The results therefore indicate the characteristic scale of the event-rate
variation associated with neutron-skin changes, without implying an
optimized experimental sensitivity.


\section{Numerical procedure for the sensitivity estimates}
\label{app:numerical_sensitivity}
This Appendix briefly summarizes the numerical procedure used to obtain the
target--dependent neutron--skin sensitivities reported in
Table~\ref{tab:deltaR_sensitivity}.
The aim is to document the concrete numerical pipeline employed in the calculations,
rather than to repeat the analytic formalism discussed in the main text.

\subsection{Binned event prediction with detector response}
For each target nucleus, we construct the recoil--energy spectrum induced by
monoenergetic KDAR neutrinos at $E_\nu=236~\mathrm{MeV}$.
For a given neutron--skin thickness $\Delta R_{np}$, the unsmeared differential
event rate per unit KDAR fluence is written as
\begin{equation}
	\frac{dN}{dT}(T;\Delta R_{np})
	=
	N_T\,
	\frac{d\sigma}{dT}(T;\Delta R_{np}) ,
	\label{eq:app_dNdT_true}
\end{equation}
where $N_T$ is the number of target nuclei corresponding to the assumed exposure.
The CE$\nu$NS differential cross section includes the full weak form factor and
is evaluated using a fixed low--energy effective weak mixing angle,
$\sin^2\theta_W(Q^2\simeq0)=0.2386$, appropriate for neutral--current processes at
momentum transfers $Q\sim\mathcal{O}(10$--$100)\,\mathrm{MeV}$.

Detector effects are incorporated through a Gaussian energy--resolution model.
Rather than performing nested numerical integrations for each recoil bin, the
smearing is implemented by constructing a detector response matrix,
$R_{ik}$, which gives the probability for an event with true recoil energy $T_k$
to be reconstructed in the observed recoil bin $[T_i,T_{i+1}]$.
The expected number of events per unit fluence in bin $i$ is then obtained as
\begin{equation}
	N_i(\Delta R_{np})
	=
	\sum_k
	R_{ik}\,
	\frac{dN}{dT}(T_k;\Delta R_{np})\,\Delta T ,
	\label{eq:app_bincount}
\end{equation}
where $\Delta T$ is the spacing of the discretized true--energy grid.
This approach avoids repeated adaptive integrations and allows for an efficient
evaluation of binned spectra over a wide range of $\Delta R_{np}$ values.

\subsection{Extraction of the $1\sigma$ sensitivity}
To quantify the statistical sensitivity to the neutron--skin thickness, binned
predictions at a trial value $\Delta R_{np}$ are compared to a reference spectrum
computed at $\Delta R_{np}^{\rm ref}$.
Assuming Poisson statistics and neglecting systematic uncertainties, we define
\begin{equation}
	\Delta \chi^2(\Delta R_{np})
	=
	\sum_i
	\frac{\left[
		N_i(\Delta R_{np})-N_i(\Delta R_{np}^{\rm ref})
		\right]^2}
	{N_i(\Delta R_{np}^{\rm ref})}.
	\label{eq:app_dchi2}
\end{equation}
Since the event counts scale linearly with the integrated KDAR fluence,
$\Delta\chi^2$ is first computed per unit fluence and subsequently rescaled to
the benchmark fluence quoted in Table~\ref{tab:deltaR_sensitivity}.
The $1\sigma$ sensitivity is obtained by solving
$\Delta\chi^2(\Delta R_{np})=1$ in the vicinity of $\Delta R_{np}^{\rm ref}$ using
a one--dimensional root search.

\subsection{Numerical settings and stability checks}
All numerical evaluations were performed in \texttt{Mathematica} using a fixed
discretization of the true recoil energy with step size $\Delta T$, chosen to
balance numerical efficiency and accuracy.
The resulting sensitivities were verified to be stable against moderate variations
of the true--energy step size and recoil--energy bin width.
The procedure described above reproduces the values reported in
Table~\ref{tab:deltaR_sensitivity} and provides a transparent and computationally
efficient framework for the sensitivity estimates presented in this work.



\end{document}